\documentclass[aps,pre,groupaddress,twocolumn]{revtex4-1}

\usepackage{graphicx} 
\usepackage{amsmath,amssymb}
\usepackage{natbib}
\usepackage{color}

\begin{document}

\title{Species exclusion and coexistence in a noisy voter model with a competition-colonization tradeoff}

\author{Ricardo Martinez-Garcia}    
\email[]{ricardom@ictp-saifr.org}
\affiliation{ ICTP-South American Institute for Fundamental Research - Instituto de F\'isica Te\'orica da UNESP, Rua Dr. Bento Teobaldo Ferraz 271, 01140-070 São Paulo, Brazil.}

\author{Crist\'obal L\'opez}
\affiliation{IFISC (CSIC-UIB), Instituto de F\'isica Interdisciplinar y Sistemas Complejos, Campus Universitat de les Illes Balears, E-07122, Palma de Mallorca, Spain. }

\author{Federico Vazquez}
\affiliation{ Instituto de C\'alculo, FCEN, Universidad de Buenos Aires and CONICET, Buenos Aires, Argentina}

\date{\today}

\begin{abstract}
  We introduce an asymmetric noisy voter model to study the joint effect of immigration and a competition-dispersal tradeoff in the dynamics of two species competing for space in regular lattices. Individuals of one species can invade a nearest-neighbor site in the lattice, while individuals of the other species are able to invade sites at any distance but are less competitive locally, i.e., they establish with a probability $g \le 1$.  The model also accounts for immigration, modeled as an external noise that may spontaneously replace an individual at a lattice site by another individual of the other species. This combination of mechanisms gives rise to a rich variety of outcomes for species competition, including exclusion of either species, mono-stable coexistence of both species at different population proportions, and bi-stable coexistence with proportions of populations that depend on the initial condition. Remarkably,  in the bi-stable phase,  the system undergoes a discontinuous transition as the intensity of immigration overcomes a threshold, leading to a half loop dynamics associated to a cusp catastrophe, which causes the irreversible loss of the species with the shortest dispersal range.
\end{abstract}

\maketitle

\section{Introduction}

Studying the balance between species competitive and dispersal abilities is fundamental to understanding the role of space in maintaining biodiversity \cite{Skellam1951, Levins-1971, Levin1974,  Hastings1980, Tilman-1994,Kneitel-2004}. Competition-dispersal (or competition-colonization) tradeoffs have primarily been invoked to explain the structure of plant communities \cite{Tilman-1990b,Bolker1999,Coomes-2003,Kisdi2003}, but have also been measured in virus \citep{Ojosnegros2010}, insects \cite{Ferzoco-2019}, bacteria \cite{Nadell-2011,Yawata-2014}, rotifers and	 protozoan \citep{Cadotte-2006}, and slime molds \cite{Tarnita-2015a,Martinez-Garcia-2016,Martinez-Garcia-2017,Rossine-2020} among other lineages. A key result is that species that are weaker competitors may persist in the community because their enhanced dispersal, either in distance or rate, allows them to colonize empty patches before the stronger competitor arrives \cite{Levins-1971}. Beyond ecological systems, competition-colonization tradeoffs have also been suggested to control tumor growth \cite{Orlando-2013}.

Different modeling approaches can be used to investigate how dispersal rates turn species coexistence stable via a competition-colonization tradeoff \cite{Levins-1971,Hanski-1997}. However, only spatial models of interacting particles, either on-lattice \citep{Harada1994, Pigolotti2018, Ellner-2001} or off-lattice \citep{Bolker1999, Martinez-Garcia2015a,Surendran2018,Surendran-2020}, allow investigation of the tradeoff between dispersal distance and competitive ability \citep{Durrett1994, Durrett-1998, Bolker1999}. Within this family of models, the voter model (VM) was introduced in \citet{Clifford-1973} as a simple model for the dynamics of two species that compete to colonize a territory, represented by a regular lattice. In the VM, each site of the lattice is occupied by an individual of either species and thus assigned a binary state. At each time step of the dynamics, one lattice site, the {\it receiver} or {\it voter} \citep{Sood-2008}, is chosen randomly and adopts the state of an {\it invader}, which is one of its four nearest-neighbors, also chosen at random. After this seminal work, the VM has been applied to other different processes, such as opinion dynamics \cite{Holley-1975,Fortunato-2007} and catalytic reactions \cite{Krapivsky-1992,Frachebourg-1996}, and studied in complex networks \cite{Sood-2008} and with continuum Langevin equations \cite{Vazquez-2008-b}. The voter model, however, does not allow for coexistence, and the only possible stationary state is the complete dominance of one species and the exclusion of the other (or \textit{consensus} in the context of social sciences). That is, an absorbing state in which all lattice site are in the same state. The invasion process is a variation of the voter model in which the {\it invader} is selected first and the {\it receiver}, second \citep{Castellano2005}. Although this different ordering in the updating rule can change the dynamics in complex networks, it is unimportant in regular lattices \citep{Sood-2008}.

A competition-dispersal tradeoff can be studied in voter-like models, including the invasion process, by considering that species differ in their dispersal range, represented by the size of the lattice neighborhood that they can potentially invade, and their competitive strength, given by the probability of displacing a non-specific resident in a given lattice site \cite{Durrett-1998}. Using numerical simulations, \citet{Durrett-1998} studied the effect of population spatial structure in determining the outcome of competition in 
 hierarchical populations (one species has a competitive advantage over all the other) and nonhierarchical populations. 
In one of the studied scenarios, they investigated the effect of tradeoffs between competition and dispersal distance in a two-species system. They showed that such tradeoff does not allow species coexistence and hence one species always excludes the other and occupies the entire territory. The identity of the excluded species will depend on the shape of the tradeoff and the initial population sizes, which indicates the existence of a bi-stability regime in the system. Moreover, \citet{Durrett-1998} showed that these patterns of species exclusions and the transitions among them are not explained by a mean field approximation, suggesting that spatial correlations in the population determine the outcome of the competitive interaction. 

Following an approach similar to \citet{Durrett-1998} but considering off-lattice simulations, \citet{Minors-2018} studied a tradeoff between interaction range and conversion strength in a VM for opinion dynamics. Their results show that opinions with larger spreading range but smaller transmission probability are more likely to spread through the population. Finally, \citet{Albano-2011} studied a VM in which the range of interactions is probabilistic, and found a multidimensional crossover behavior, from one-dimension to infinite dimensions or mean-field, as the probability of long-range interactions increases. In none of these studies the authors found stable coexistence of the competing species (or opinions).

Another extended version of the VM is the so-called \textit{noisy voter model} (NVM; see \cite{Granovsky-1995}  and references therein), which incorporates the possibility of spontaneous changes  in the state of lattice sites. The NVM was introduced by \citet{Kirman-1993} to model the stochastic recruitment behavior of groups of ants that suddenly switch their attention between two food sources \cite{Pasteels-1987}. Other early works investigated the dynamics of catalytic reactions \cite{Fichthorn-1989,Considine-1989,Clement-1991} and, more recently, the dynamics of the NVM has also been studied in complex networks \cite{Carro-2016,Peralta-2018,Khalil-2019}. A multi-state NVM has  been proposed to explain the emergence of flocking out from pairwise stochastic interactions \cite{Baglietto-2018,Vazquez-2019,Loscar-2021}. Experimental evidence of these noisy voter-like interactions was found recently in groups of fish, where schooling is induced by the intrinsic noise that arises from the finite number of interacting individuals \cite{Jhawar-2020}.  In all these cases,  the external noise eliminates the absorbing (consensus) states and,  in the two-state NVM,  the noise intensity induces a transition from a bi-stable phase characterized by a bimodal stationary distribution of opinion density to a mono-stable phase with a unimodal distribution \citep{Considine-1989}.  In the multistate NVM,  this noise-driven transition happens between regimes with multimodal and unimodal stationary distributions \citep{Herrerias-Azcue2019}.  In both cases,  in the mono-stable phase,  the system fluctuates around a state in which both opinions are equally represented. 

Here, we introduce and analyze a novel NVM in which each of the competing species can invade lattice sites within different ranges. The motivation of the model is in the context of species interactions and competition-dispersal tradeoffs. Therefore, longer dispersal distances are penalized with lower probability of displacing non-specific individuals from a lattice site. Using this model, we study whether and in which conditions a competition-dispersal tradeoff and external noise (mimicking immigration) may give rise to species coexistence. To address this question, we  develop a pair approximation  and perform Monte Carlo simulations of the model that   unveil the existence of various dynamical regimes. We find that both mono-stable and bi-stable  coexistence of species is possible if the competition-dispersal  tradeoff and immigration act simultaneously, and determine the conditions and parameter regimes that lead to each of  these scenarios. As a byproduct, we show that in the limit of no immigration --the original model studied  in \cite{Durrett-1998}-- the pair approximation provides a qualitative description of the transition between species dominance.

The outline of the paper is the following. The model and its dynamics are defined in Section~\ref{model}.  In Section~\ref{mean-field} we develop the pair approximation approach and study the cases with and without immigration separately. Monte Carlo results in both one and two-dimensional lattices are given in Section~\ref{Monte-Carlo}, and we summarize our results and provide some conclusions in Section~\ref{discussion}.

\section{The model}
\label{model}

We consider an invasion process in which two species, labeled by $C$ and $D$, compete for a territory represented by a one-dimensional lattice of $N$ sites and periodic boundary conditions. We will extend our results to two-dimensinoal lattices in section \ref{sec:MC2D}. Because the dynamics runs on a regular lattice, it is equivalent to a voter updating rule \citep{Sood-2008}.  Each of the $N$ sites is occupied by one individual of either species $C$ or $D$. Species differ on how they balance a competition-dispersal tradeoff. Individuals from species $C$ ($C$ stands for competitors) can only colonize one of their two nearest-neighbor sites, whereas individuals from species $D$ ($D$ stands for dispersers), can colonize any other site of the lattice. To account for the cost of enhanced dispersal range typical of competition-colonization tradeoffs, dispersers have a reduced local competition strength. $D$-replicates have a probability $g_{\mbox{\tiny{D}}}=g\leq 1$ of  replacing resident individuals after dispersal, while competitors  establish with probability $g_{\mbox{\tiny{C}}}=1$. The macroscopic state of the system is determined by the fraction of sites occupied by competitors and dispersers, $\rho_{\mbox{\tiny{C}}}$ and $\rho_{\mbox{\tiny{D}}}$, respectively, with $\rho_{\mbox{\tiny{D}}}(t) + \rho_{\mbox{\tiny{C}}}(t) = 1$ for all times $t \ge 0$.

\begin{figure}
  \includegraphics[width=0.43\textwidth]{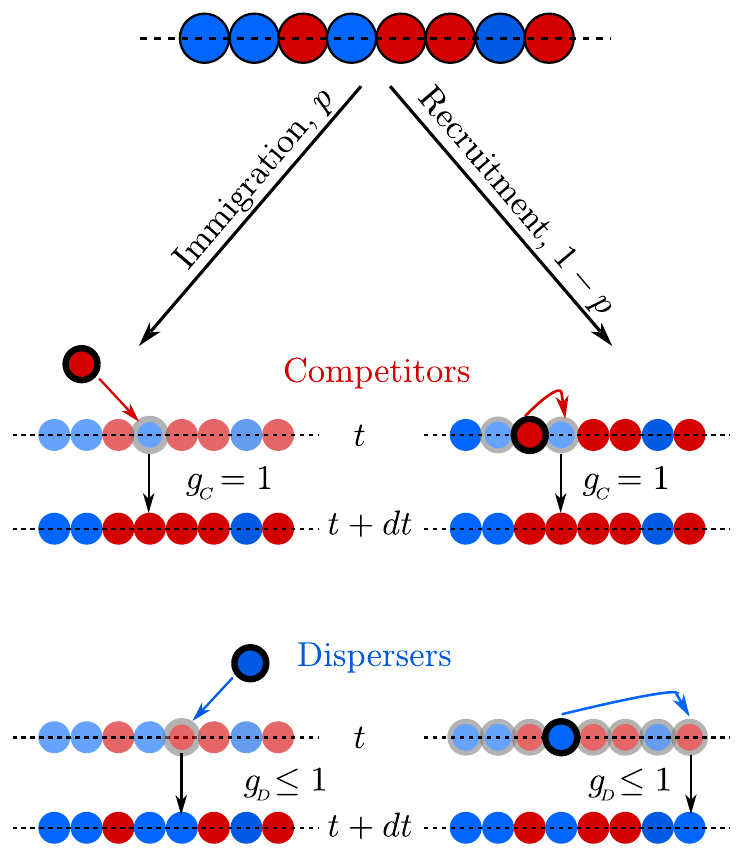}
  \caption{Model schematic. Each site of a one-dimensional  regular lattice is occupied by one individual that belongs to either species $C$ (red circles) or $D$ (blue circles). At each time step, one of two possible events takes place.  With probability $p$, one individual is randomly chosen and replaced by an individual of the other species with establishing probabilities $g_{\mbox{\tiny{C}}}=1$ and $g_{\mbox{\tiny{D}}}=g$ for $C$ and $D$ species, respectively (immigration).  With the complementary probability $1-p$, one  randomly chosen individual reproduces and, with the species-dependent establishing probability defined for immigration, its offspring invades a site chosen at random within its dispersal range}
\label{fig:model}
\end{figure}

In a single time step $dt=1/N$ of the dynamics, one of the two following processes takes place (see Fig.~\ref{fig:model}):
\begin{itemize}
 \item \emph{Immigration}.  With probability $p$, one individual is chosen at random and replaced by an individual from the other species with species-dependent probability.  A disperser is replaced by a competitor with probability one, while a competitor is replaced by a disperser with \textit{{establishment}} probability $g$.  This updating rule, which results in an external noise, represents the immigration of individuals from other patches.
   
\item \emph{Recruitment}.  With the complementary probability $1-p$, one randomly chosen individual replicates.  Following replication, this individual can potentially invade a site chosen at random within its species-specific neighborhood (dispersal range).  A competitor invades a nearest-neighbor (NN) site with probability $1$, while a disperser invades any other site of the lattice with probability $g$.
\end{itemize}

Note that the model accounts for two different dispersal processes that act independently of each other: inter-patch dispersal, represented by immigration, and intra-patch dispersal following individual reproduction. We assume that each of these processes is mediated by different mechanisms and therefore immigration is independent of the competition-colonization tradeoff. Specifically, we consider that immigration is mediated by external factors that are species-independent. Therefore, both dispersers and competitors arrive at the patch at the same rate and the external noise that models immigration is symmetric. Conversely, we assume that the colonization-competition tradeoff is determined by physiological species-dependent properties of the individual, such as seed size or spore size \cite{Coomes-2003, Martinez-Garcia-2017}

Given this set of ingredients, we investigate the competition between  a nearest-neighbor and a mean-field  dispersal strategy  in which enhanced dispersal comes at the cost of a lower competitive ability at the local scale. The combination of the external noise introduced by immigration,  and the differences in dispersal ranges and site colonization probabilities between species constitutes the novel ingredient of our model with respect to previous versions of the voter model or invasion process, and posterior applications in biological dynamics \cite{Durrett-1998}.

\section{Pair Approximation analysis of the model}
\label{mean-field}

In this section we develop an analytical approach that captures the most salient features of the behavior of the model. This approach is based on a \textit{pair approximation} (PA), a moment closure approximation that takes into account the spatial correlations between the states of individuals that are first neighbors in the lattice \cite{Bolker1999,Ellner-2001,Iwasa-2010,Surendran-2020,Vieira-2020}, similar to that used in \cite{Demirel-2014}.  We start by analyzing the case in which there is only recruitment, and thus the dynamics is reduced to that of two species, $C$ and $D$, which compete for territory by simple invasion. We then incorporate the external noise into the analysis to explore how immigration affects the steady state of the system.

\subsection{Only recruitment case: $p=0$}
\label{recruitment}

{\it Derivation of Pair Approximation equations.-} When there is no immigration ($p=0$), the only control parameter is the cost of long-range dispersal, represented by the establishment probability $g$ of dispersers.  In this limit, our model reduces to that proposed by \citet{Durrett-1998}. Using a PA, we extend their analysis and provide an analytical support to the results observed in numerical simulations. First, we bound the range of values of the establishment probability for which the model is bi-stable. Second, within the bi-stable phase, we provide an estimation for the relation between initial competitor density and establishment probability at which the system undergoes a transition from competitor to disperser dominance.

We consider that the fraction of sites occupied by competitors ($C$--sites),  $\rho_{\mbox{\tiny{C}}}$, evolves in the $N \to \infty$ limit according to the following rate equation:
\begin{equation}
  \frac{d \rho_{\mbox{\tiny{C}}}}{dt} = W^+(\rho_{\mbox{\tiny{C}}}) - W^-(\rho_{\mbox{\tiny{C}}}),
  \label{drhoSdtmaestra}
\end{equation}
where $W^+(\rho_{\mbox{\tiny{C}}})=W(\rho_{\mbox{\tiny{C}}} \to \rho_{\mbox{\tiny{C}}}+1/N)$ and $W^-(\rho_{\mbox{\tiny{C}}})=W(\rho_{\mbox{\tiny{C}}} \to \rho_{\mbox{\tiny{C}}}-1/N)$ are, respectively, the gain and loss transition probabilities in a time step $dt=1/N$.  These probabilities can be estimated using a PA in which correlations to second nearest-neighbors are neglected. The transition $W^-$ is calculated as 
\begin{eqnarray}
  W^-(\rho_{\mbox{\tiny{C}}}) = g \, \rho_{\mbox{\tiny{C}}} (1-\rho_{\mbox{\tiny{C}}}).
  \label{W-}
\end{eqnarray}
That is, in a single time step a $D$--site is selected at random for replication with probability $P(D)=\rho_{\mbox{\tiny{D}}}=1-\rho_{\mbox{\tiny{C}}}$.  Then, a random $C$--site is chosen with probability $\rho_{\mbox{\tiny{C}}}$, which changes to state $D$ with probability $g$ ($D$--offspring establishment).  Analogously, $W^+$ is estimated as
\begin{eqnarray}
  W^+(\rho_{\mbox{\tiny{C}}}) = \rho_{\mbox{\tiny{C}}} \, P(D|C) = \rho_{\mbox{\tiny{CD}}},
  \label{W+}
\end{eqnarray}
where $\rho_{\mbox{\tiny{CD}}}$ is the fraction of neighboring $CD$ pairs, i.e., the fraction of NN sites occupied by a competitor and a disperser.  That is, with probability $P(C)=\rho_{\mbox{\tiny{C}}}$ a $C$--site is randomly chosen, which then invades a NN site that is in state $D$ with probability $P(D|C)$, where $P(D|C)=P(CD)/P(C)=\rho_{\mbox{\tiny{CD}}}/\rho_{\mbox{\tiny{C}}}$ is the conditional probability that a $D$--site is a NN of a $C$--site. 

Replacing Eqs.~(\ref{W-}) and (\ref{W+}) into Eq.~(\ref{drhoSdtmaestra}), we arrive at
\begin{equation}
  \frac{d \rho_{\mbox{\tiny{C}}}}{dt} = \rho_{\mbox{\tiny{CD}}} - g \, \rho_{\mbox{\tiny{C}}} (1-\rho_{\mbox{\tiny{C}}}).
  \label{drhoSdt}
\end{equation}

As the equation for $\rho_{\mbox{\tiny{C}}}$ depends on the density of $CD$ pairs, we next have to derive an equation for $\rho_{\mbox{\tiny{CD}}}$ to obtain a closed system. For this calculation we need  the complete set of conditional probabilities 
\begin{subequations}
  \begin{alignat}{4}
    P(C|D) &= \frac{\rho_{\mbox{\tiny{CD}}}}{\rho_{\mbox{\tiny{D}}}}, \\
    P(C|C) &= \frac{\rho_{\mbox{\tiny{CC}}}}{\rho_{\mbox{\tiny{C}}}} = 1 - \frac{\rho_{\mbox{\tiny{CD}}}}{\rho_{\mbox{\tiny{C}}}}, \\
    P(D|C) &= \frac{\rho_{\mbox{\tiny{CD}}}}{\rho_{\mbox{\tiny{C}}}}, \\
    P(D|D) &= \frac{\rho_{\mbox{\tiny{DD}}}}{\rho_{\mbox{\tiny{D}}}} = 1 - \frac{\rho_{\mbox{\tiny{CD}}}}{\rho_{\mbox{\tiny{D}}}},  
  \end{alignat}
  \label{conditional}
\end{subequations}    
where we have used the relations 
\begin{eqnarray}
 \rho_{\mbox{\tiny{CD}}}+\rho_{\mbox{\tiny{CC}}}=\rho_{\mbox{\tiny{C}}}~~~ \mbox{and} ~~~ \rho_{\mbox{\tiny{CD}}}+\rho_{\mbox{\tiny{DD}}}=\rho_{\mbox{\tiny{D}}}
  \label{relations}
\end{eqnarray}
that reflect the conservation of the fraction of pairs and sites.  The rate equation for $\rho_{\mbox{\tiny{CD}}}$ reads
\begin{equation}
  \frac{d \rho_{\mbox{\tiny{CD}}}}{dt} = 2 \, W^+(\rho_{\mbox{\tiny{CD}}}) - 2 \, W^-(\rho_{\mbox{\tiny{CD}}}),
  \label{drhoSLdt-0}
\end{equation}
where the prefactor $2$ comes from the change in $\rho_{\mbox{\tiny{CD}}}$ by $2/N$ in a single update.  To calculate the gain and loss transition probabilities $W^+(\rho_{\mbox{\tiny{CD}}})=W(\rho_{\mbox{\tiny{CD}}} \to \rho_{\mbox{\tiny{CD}}}+2/N)$ and $W^-(\rho_{\mbox{\tiny{CD}}})=W(\rho_{\mbox{\tiny{CD}}} \to\rho_{\mbox{\tiny{CD}}}-2/N)$, respectively, we consider the possible transitions that lead to a change in the density of $CD$ pairs.  For instance, if $\sigma$ is a discrete variable that denotes the state of a lattice site, $\sigma_i = \lbrace C,D\rbrace$, the transition $CDC \to CCC$ takes place when a randomly chosen site $i$ in state $\sigma_i=C$ invades a NN site $i+1$ in state $\sigma_{i+1}=D$, whose other neighboring site $i+2$ is in state $\sigma_{i+2}=C$. This happens with probability $P(CDC \to CCC) = P(C)P(D|C)P(C|DC)$.  Within a PA, the probability $P(C|DC)$ that $\sigma_{i+2}=C$ given that $\sigma_{i+1}=D$ and $\sigma_{i}=C$ can be approximated as $P(C|D)$ if we neglect correlations between $\sigma_{i+2}$ and $\sigma_{i}$. Therefore, using Eqs.~(\ref{conditional}) we have $P(CDC \to CCC) \simeq \rho_{\mbox{\tiny{CD}}}^2/\rho_{\mbox{\tiny{D}}}$.  Calculating in the same way all possible transition probabilities that lead to a change in $\rho_{\mbox{\tiny{CD}}}$ in a time step, we obtain
\begin{subequations}
  \begin{alignat}{2}
    W^+(\rho_{\mbox{\tiny{CD}}}) &= P(CCC \to CDC),  \\
    W^-(\rho_{\mbox{\tiny{CD}}}) &= P(DCD \to DDD) + P(CDC \to CCC), 
  \end{alignat}
  \label{W-W+}  
\end{subequations}
where
\begin{subequations}
  \begin{alignat}{2}
  \label{SSS-SLS}
  P(CCC \to CDC) &\simeq \frac{g \, (1-\rho_{\mbox{\tiny{C}}})(\rho_{\mbox{\tiny{C}}}-\rho_{\mbox{\tiny{CD}}})^2}{\rho_{\mbox{\tiny{C}}}}, \\
  \label{LSL-LLL}
  P(DCD \to DDD) &\simeq \frac{g \, (1-\rho_{\mbox{\tiny{C}}})\rho_{\mbox{\tiny{CD}}}^2}{\rho_{\mbox{\tiny{C}}}}, \\
  \label{SLS-SSS}
  P(CDC \to CCC) &\simeq \frac{\rho_{\mbox{\tiny{CD}}}^2}{1-\rho_{\mbox{\tiny{C}}}}.
  \end{alignat}
  \label{P-trans}
\end{subequations}
The first two probabilities, Eqs.~(\ref{SSS-SLS}) and (\ref{LSL-LLL}), correspond to the transition $C \to D$ due to the long-range invasion of a disperser, which happens with probability $g$.  The other possible transitions $CDD \to CCD$ and $CCD \to CDD$ do not generate a change in $\rho_{\mbox{\tiny{CD}}}$, and thus they are omitted.  

Finally, combining Eqs.~(\ref{drhoSLdt-0})-(\ref{P-trans}) we obtain the following equation for the evolution of $\rho_{\mbox{\tiny{CD}}}$:
\begin{equation}
  \frac{d \rho_{\mbox{\tiny{CD}}}}{dt} = 2 g (1-\rho_{\mbox{\tiny{C}}})(\rho_{\mbox{\tiny{C}}}- 2 \rho_{\mbox{\tiny{CD}}})-\frac{2\rho_{\mbox{\tiny{CD}}}^2}{1-\rho_{\mbox{\tiny{C}}}}.
  \label{drhoSLdt}
\end{equation}
The closed system of Eqs.~(\ref{drhoSdt}) and (\ref{drhoSLdt}) for $\rho_{\mbox{\tiny C}}$ and $\rho_{\mbox{\tiny{CD}}}$ represents an approximate macroscopic description of the evolution of the system, which we analyze below.

\begin{figure}
  \includegraphics[width=0.4\textwidth]{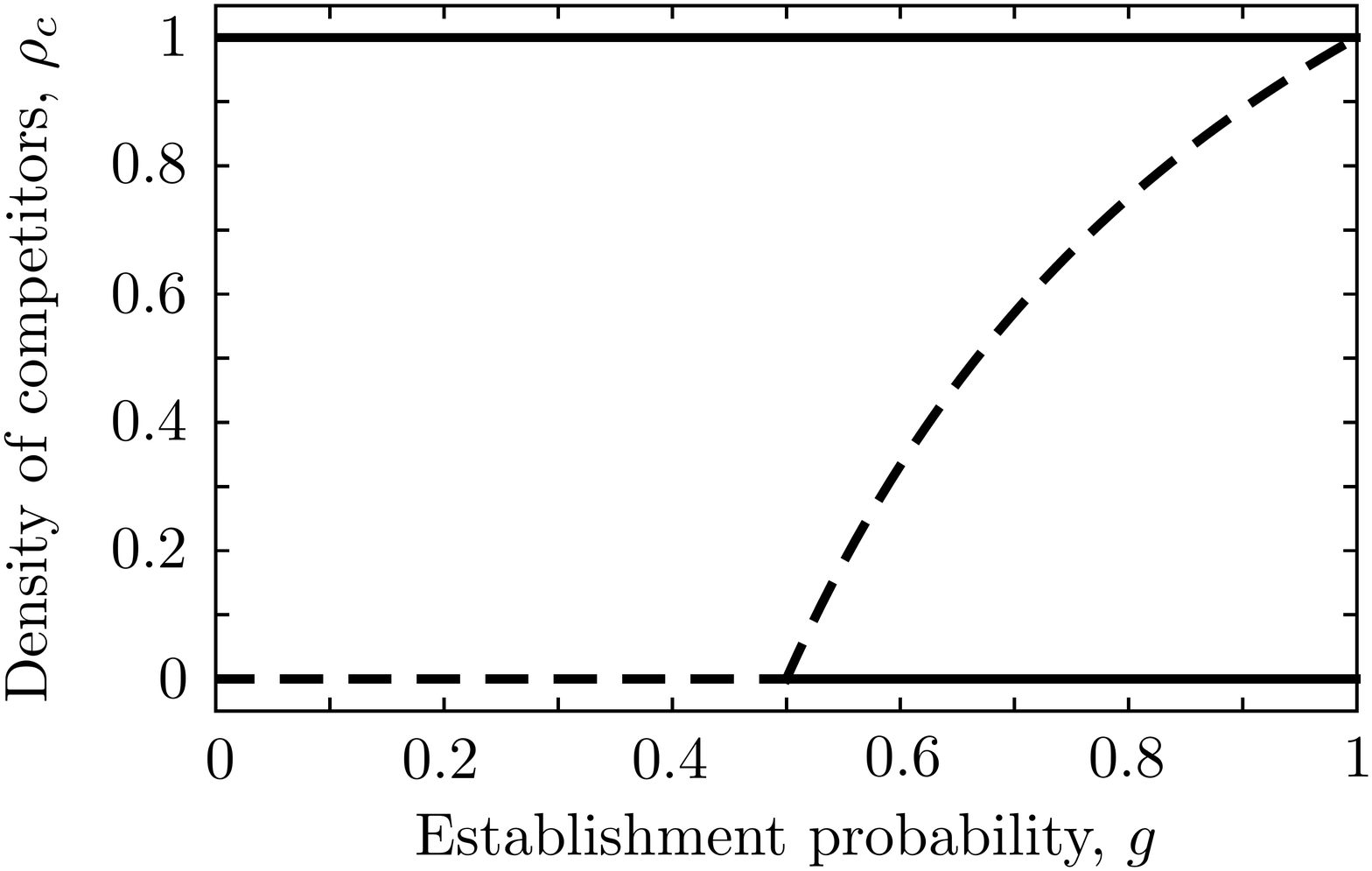}
  \caption{Bifurcation diagram of the model predicted by the pair approximation with $p=0$. A PA analysis predicts a region of $C$-dominance for $0\leq g<1/2$ and a bi-stability region for  $1/2 < g\leq 1$. Solid and dashed lines indicate the stable and unstable fixed points, respectively.}
  \label{fig:nop-bifurcation}
\end{figure}

\begin{figure}
  \includegraphics[width=0.35\textwidth]{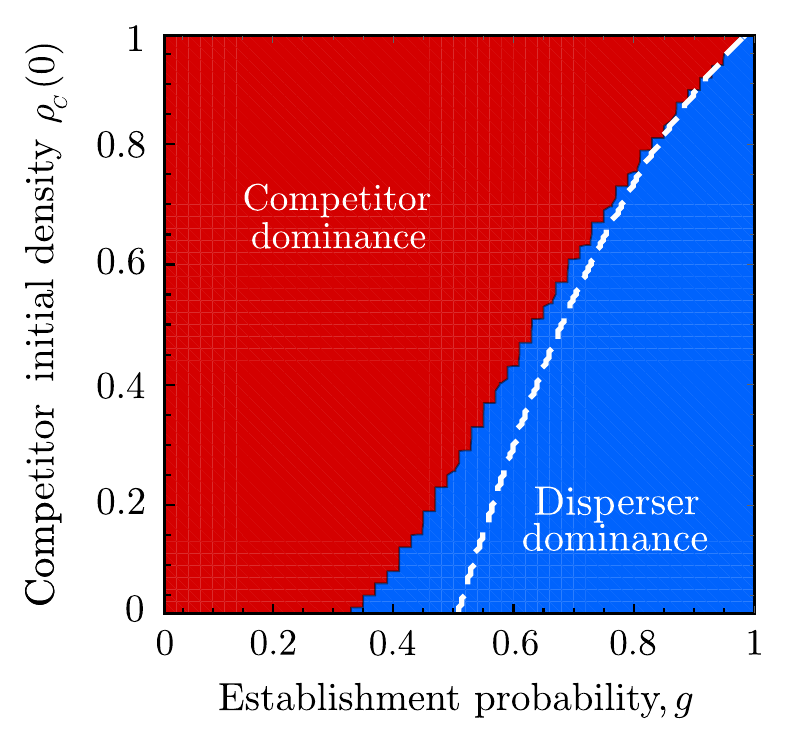}
  \caption{Phase diagram of the model without immigration $(p=0)$. Color background represents the stationary state in the MC simulations of the stochastic dynamics. Competitors dominate in the red region and dispersers dominate in the blue region. The white dashed line indicates the prediction of the pair approximation for the transition from competitor dominance to disperser dominance. We conducted the MC simulations on a one-dimensional lattice with $N=10^5$ and ran $100$ independent realizations for each $(\rho_{\mbox{\tiny{C}}}(0), g)$ pair.}
  \label{fig:phase}
\end{figure}

\begin{figure}
  \includegraphics[width=0.45\textwidth]{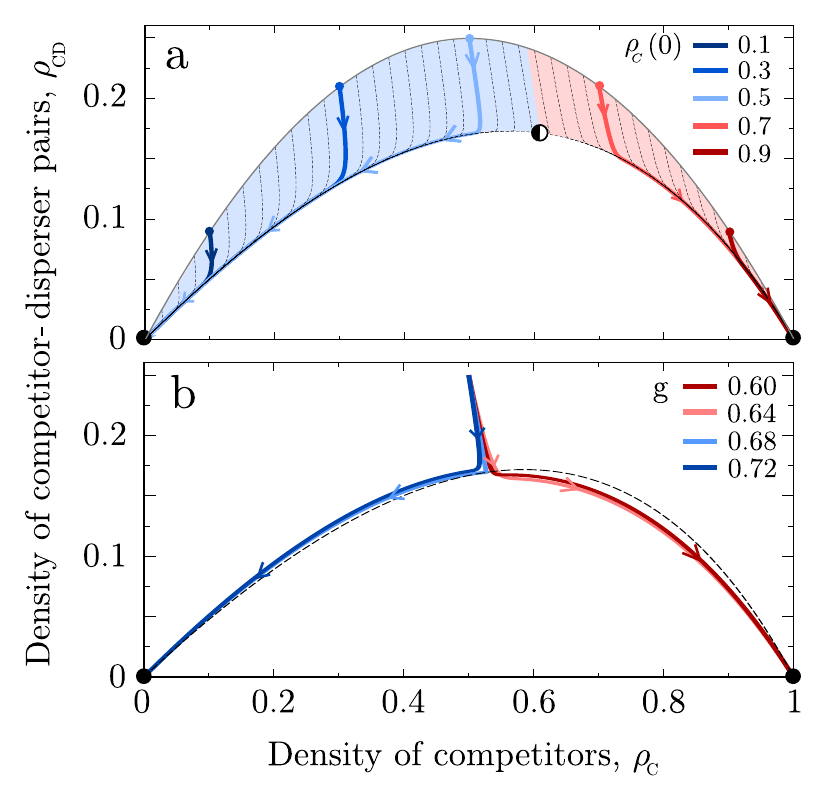}
  \caption{Schematic flow diagram of the system in the $\rho_{\mbox{\tiny{C}}}-\rho_{\mbox{\tiny{CD}}}$ plane. The stable fixed points $(0,0)$ and $(1,0)$ denoted by black disks represent the dominance of $D$ and $C$ species, respectively. a) The establishment probability is kept constant at $g=0.72$ (bi-stable phase) and the initial condition varied along the curve $\rho_{\mbox{\tiny{CD}}} = \rho_{\mbox{\tiny{C}}}(1-\rho_{\mbox{\tiny{C}}})$ (gray curve). The semi-filled disk indicates the position of the saddle node for the value of $g$ used in the panel. b) The initial condition is kept constant at $(\rho_{\mbox{\tiny{C}}}(0)=1/2, \rho_{\mbox{\tiny{CD}}}(0)=1/4)$ and the establishment probability $g$ varies in a way that trajectories change the ending point at the transition value $g_{\mbox{\tiny{T}}} \simeq 0.68$ from competitor to disperser dominance.  The black dashed curve given by $\rho_{\mbox{\tiny{CD}}}=\rho_{\mbox{\tiny{C}}}(1-\rho_{\mbox{\tiny{C}}})/(2-\rho_{\mbox{\tiny{C}}})$ corresponds to the line of all  coexistence saddle points $\vec{\rho_{co}}^*$ for values of $g$ in $[0,1]$. }
  \label{diagram}
\end{figure}

{\it Analysis of Pair approximation Equations.-} We are interested in the behavior of the stationary value of $\rho_{\mbox{\tiny{C}}}$ as the germination probability $g$ is changed. The system of coupled equations (\ref{drhoSdt}) and (\ref{drhoSLdt}) has two trivial fixed points $(\rho_{\mbox{\tiny{C}}}^*,\rho_{\mbox{\tiny{CD}}}^*)=(0,0)$ and $(1,0)$, corresponding to the complete dominance of  dispersers and competitors, respectively, and a third non-trivial fixed point
\begin{equation}
  \vec{\rho_{co}}^* = \left(\frac{2g-1}{g}, \frac{(1-g)(2g-1)}{g}\right)
\end{equation}
that represents a coexistence of both species. Note that $ \vec{\rho_{co}}^*$ has a physical meaning ($0 \leq \rho_{\mbox{\tiny{C}}}^* \leq 1$ and $0 \leq \rho_{\mbox{\tiny{CD}}}^* \leq 1$) only for $1/2 \leq g \leq 1$, while $ \vec{\rho_{co}}^*$ lays on the negative quadrant $\rho_{\mbox{\tiny{C}}}^*<0$ and $\rho_{\mbox{\tiny{CD}}}^*<0$ for $0\leq g <1/2$. Starting from a given initial condition $\vec{\rho_0}=(\rho_{\mbox{\tiny{C}}}(0),\rho_{\mbox{\tiny{CD}}}(0))$, the system evolves towards one of the three fixed points whose basin of attraction contains the initial point $\vec{\rho_0}$.  In Appendix~\ref{stability} we perform a linear stability analysis that leads to the following picture (Figs.~\ref{fig:nop-bifurcation} and \ref{fig:phase}):  

\begin{enumerate}
\item For $0 \le g <1/2$, as discussed above, the coexistence point $\vec{\rho_{co}}^*$ is a non-physical state. The fixed point $(1,0)$ is stable and $(0,0)$ is a saddle point. Starting from any physical situation ($\rho_{\mbox{\tiny{C}}}(0)>0$ and $\rho_{\mbox{\tiny{CD}}}(0)>0$) the system ends in the $(1,0)$ fixed point corresponding to a $C$--dominance (Fig.~\ref{fig:nop-bifurcation})

\item For $1/2 < g \le 1$ the fixed  points $(0,0)$ and $(1,0)$ are stable and $\vec{\rho_{co}}^*$ is a saddle point (Fig.~\ref{fig:nop-bifurcation}). Therefore, the  system is in a bi-stable phase where, depending on the initial  condition and $g$, the final state is  either the one of competitor $(1,0)$ or disperser $(0,0)$ dominance (Fig.~\ref{fig:nop-bifurcation}). That is, the system falls into $(1,0)$ for $g < g_{\mbox{\tiny{T}}}$ and into $(0,0)$ for $g > g_{\mbox{\tiny{T}}}$, where $g_{\mbox{\tiny{T}}} = g_{\mbox{\tiny{T}}}(\vec{\rho_0})$ is a transition point that depends on the initial state $\vec{\rho_0}$.  In Fig.~\ref{fig:phase} we plot the transition line in the  $g$--$\rho_{\mbox{\tiny{C}}}(0)$ phase diagram for an  initial condition  $\vec{\rho_0}=[\rho_{\mbox{\tiny{C}}}(0),\rho_{\mbox{\tiny{C}}}(0) \,  \rho_{\mbox{\tiny{D}}}(0)]$ that corresponds to a density  $\rho_{\mbox{\tiny{C}}}(0)$ of individuals uniformly distributed  over the lattice.  For instance, for an initial condition that  corresponds to uniform densities  $\rho_{\mbox{\tiny{C}}}(0)=\rho_{\mbox{\tiny{D}}}(0)=1/2$ ($\rho_{\mbox{\tiny{CD}}}(0)=1/4$) we found $g_{\mbox{\tiny{T}}} \simeq  0.68$ by integrating Eqs.~(\ref{drhoSdt}) and (\ref{drhoSLdt})  numerically. The trajectories of the system for various values of  $g$ are depicted in the $\rho_{\mbox{\tiny{C}}}$--$\rho_{\mbox{\tiny{CD}}}$ flow diagram of Fig.~\ref{diagram}.  We can see that the final point of the  trajectory starting at $(1/2,1/4)$ changes at $g_{\mbox{\tiny{T}}}$ (panel b).
\end{enumerate}

In summary, the analytical approach developed in this section predicts that the stable coexistence of species is not possible for any value of $g$ when there is only recruitment ($p=0$), and thus either competitors or dispersers dominate the space in the final stationary state.  Specifically, for $g$ in the range $[0,1/2)$ competitors dominate for all initial densities $\rho_{\mbox{\tiny{C}}}(0)>0$, while for $g$ in $(1/2,1]$ the model is bi-stable and the final state depends on the initial condition $\vec{\rho_0}$ \cite{Durrett-1998}. That is, either competitors dominate for $g<g_{\mbox{\tiny{T}}}(\vec{\rho_0})$ (left side of the white dashed line in Fig.~\ref{fig:phase}) or dispersers dominate for $g>g_{\mbox{\tiny{T}}}(\vec{\rho_0})$ (right side of the white dashed line).

\subsection{Recruitment and immigration: $p>0$}
\label{immigration}

We now study the model when, besides recruitment, immigration from other patches is taken into account ($p>0$).  Immigration can be implemented in our model \emph{via} an external noise that spontaneously switches the identity of the species that occupies a lattice site (note the similarity with the noisy voter model). We concluded  from the previous section that recruitment acting alone  leads to the dominance of one of the species, either  competitors or dispersers, for all values of $g$ and physical initial conditions. In this section, we study how these stable states are affected by the external noise of amplitude $p$.

{\it Pair Approximation equations.-}
The rate equations for this case can be obtained by following the same approach introduced in Section~\ref{recruitment}.  We obtain
\begin{subequations}
  \begin{alignat}{2}
    \label{drhoSdt-1}
    \frac{d \rho_{\mbox{\tiny{C}}}}{dt} &= (1-p) \left[\rho_{\mbox{\tiny{CD}}} - g \,
      \rho_{\mbox{\tiny{C}}}(1-\rho_{\mbox{\tiny{C}}})\right] + p \left[ 1-(1+g)\rho_{\mbox{\tiny{C}}} \right], \\
    \label{drhoSLdt-1}
    \frac{d \rho_{\mbox{\tiny{CD}}}}{dt} &= 2(1-p) \left[g(1-\rho_{\mbox{\tiny{C}}})(\rho_{\mbox{\tiny{C}}}-2\rho_{\mbox{\tiny{CD}}}) - \frac{\rho_{\mbox{\tiny{CD}}}^2}{1-\rho_{\mbox{\tiny{C}}}}\right]+\\
    & + 2p \left[1-(1-g) \rho_{\mbox{\tiny{C}}}-2(1+g) \rho_{\mbox{\tiny{CD}}} \right].\nonumber
  \end{alignat}
  \label{rhoS-rhoSL}
\end{subequations}   
The first terms in Eqs.~(\ref{rhoS-rhoSL}) with the prefactor $1-p$ correspond to the changes in the densities due to a recruitment event, already derived in section~\ref{recruitment}.  The second terms, with the prefactor $p$, represent a change due to immigration, as we describe next.  The gain (loss) transition probability for $\rho_{\mbox{\tiny{C}}}$ is $W^+(\rho_{\mbox{\tiny{C}}})= (1-\rho_{\mbox{\tiny{C}}})$ [$W^-(\rho_{\mbox{\tiny{C}}})=g \, \rho_{\mbox{\tiny{C}}}$], which corresponds to selecting a $D$--site ($C$--site) at random and switching its state with probability $1$ ($g$).  Then, the net change $W^+(\rho_{\mbox{\tiny{C}}})-W^-(\rho_{\mbox{\tiny{C}}})$ leads to the second term of Eq.~(\ref{drhoSdt-1}). 

For $\rho_{\mbox{\tiny{CD}}}$, the transition probabilities are  
\begin{subequations}
  \begin{alignat}{2}
    W^+(\rho_{\mbox{\tiny{CD}}}) &= P(CCC \to CDC) + P(DDD \to DCD),  \\
    W^-(\rho_{\mbox{\tiny{CD}}}) &= P(DCD \to DDD) + P(CDC \to CCC), 
  \end{alignat}
  \label{W-W+-1}  
\end{subequations}
with
\begin{subequations}
  \begin{alignat}{2}
  \label{SSS-SLSp}
  P(CCC \to CDC) & \simeq
  \frac{g \, (\rho_{\mbox{\tiny{C}}}-\rho_{\mbox{\tiny{CD}}})^2}{\rho_{\mbox{\tiny{C}}}}, \\
  \label{LLL-LSLp}
  P(DDD \to DCD) & \simeq \frac{(1-\rho_{\mbox{\tiny{C}}}-\rho_{\mbox{\tiny{CD}}})^2}{1-\rho_{\mbox{\tiny{C}}}}, \\
  \label{LSL-LLLp}
  P(DCD \to DDD) &\simeq \frac{g\, \rho_{\mbox{\tiny{CD}}}^2}{\rho_{\mbox{\tiny{C}}}}, \\
  \label{SLS-SSSp}
  P(CDC\to CCC) &\simeq \frac{\rho_{\mbox{\tiny{CD}}}^2}{1-\rho_{\mbox{\tiny{C}}}}.
  \end{alignat}
  \label{P-trans-1}
\end{subequations}
Then, the resulting net change $2 \, W^+(\rho_{\mbox{\tiny{CD}}})- 2 \, W^-(\rho_{\mbox{\tiny{CD}}})$ calculated from Eqs.~(\ref{W-W+-1}) and (\ref{P-trans-1}) gives, after doing some algebra, the second term of Eq.~(\ref{drhoSLdt-1}). 

{\it Analysis of Pair approximation Equations.-} We now analyze the steady state of the system in the entire $p-g$ space using Eqs.~(\ref{rhoS-rhoSL}).  In Appendix~\ref{fixed-points} we show that Eqs.~(\ref{rhoS-rhoSL}) have four fixed points, whose analytical expressions are hard to obtain. Instead, we estimated the fixed points by finding numerically the roots of a polynomial of degree four, within an error of $\sim 10^{-7}$.  Because we are interested in physical density values, we only focus on $\rho_{\mbox{\tiny{C}}}^*$ in the interval $[0,1]$.  Results are summarized in the phase diagram of Fig.~\ref{phase-diag}(a) that we describe next:  

\begin{enumerate}
\item For $0 \le g \le 1/2$ there is only one coexistence  fixed point in $[0,1]$,  $\rho_{\mbox{\tiny{C}}}^*$, which is stable for $0\le p \le 1$.  The stationary density of competitors in this so-called  \emph{mono-stable} phase decays as $p$ increases, from $\rho_{\mbox{\tiny{C}}}^*=1$ for $p=0$ (showed in section~\ref{recruitment}, and corresponding to survival only of competitors), to $\rho_{\mbox{\tiny{C}}}^*=1/(1+g)$ for $p=1$ [from Eq.~(\ref{drhoSdt-1})]. Thus, for any $p>0$ we have a situation with one single stable coexistence of the two species. We note that in this coexistence regime, competitors are always more abundant than dispersers for any $p$ and $g$. 

\begin{figure*}
  \includegraphics[width=\textwidth]{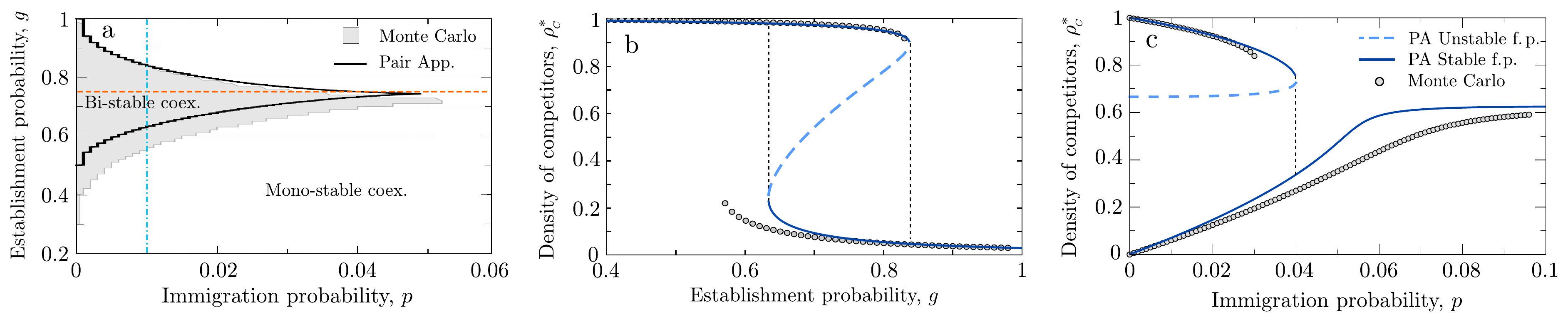}
  \caption{(a) Phase diagram on the $p-g$ space showing the region of bi-stable coexistence obtained from the pair approximation  Eqs.~(\ref{rhoS-rhoSL}) (black solid line) and Monte Carlo simulations (gray region). The dotted-dashed-cyan and dashed-orange lines indicate  the immigration probability $p=0.01$ and the establishment probability $g=0.75$ studied in detail in panel (b) and (c), respectively. (b) Stationary density of competitors $\rho_{\mbox{\tiny{C}}}^*$ vs $g$ for $p=0.01$. The system exhibits a classical hysteresis loop. (c) Stationary density of competitors $\rho_{\mbox{\tiny{C}}}^*$ vs $p$ for $g=0.75$.  The system exhibits a discontinuous transition at $p_c$, from a bi-stable phase to a mono-stable phase. In (b) and (c), solid curves correspond to PA stable fixed points; dashed curves, to PA unstable fixed points; and circles are MC simulation results.}  
  \label{phase-diag}
\end{figure*} 

\item For $1/2 < g \le 1$ the most relevant feature is the appearance of a small region of bi-stability for small values of $p$, where there are three fixed points [see phase diagram, or stability diagram \citep{Strogatz2001}, in Fig.~\ref{phase-diag}(a)]: two stable, $\rho_{\mbox{\tiny{C}}}^1$  and $\rho_{\mbox{\tiny{C}}}^3$, and one unstable,  $\rho_{\mbox{\tiny{C}}}^2$, with $\rho_{\mbox{\tiny{C}}}^1 < \rho_{\mbox{\tiny{C}}}^2 < \rho_{\mbox{\tiny{C}}}^3$.  This corresponds to a situation with two different stable coexistences $\rho_{\mbox{\tiny{C}}}^{3} > \rho_{\mbox{\tiny{C}}}^{1}$ (i.e., $\rho_{\mbox{\tiny{C}}}^{1}$ is a stable coexistence with more dispersers than competitors, and $\rho_{\mbox{\tiny{C}}}^{3}$ is the opposite situation). All along the upper (lower) boundary of the bi-stability region [black lines in Fig.~\ref{phase-diag}(a)], the model has saddle-node bifurcations in which one of the stable fixed points  $\rho_{\mbox{\tiny{C}}}^3(p)$ [$\rho_{\mbox{\tiny{C}}}^1(p)$] merges with the unstable fixed point $\rho_{\mbox{\tiny{C}}}^2(p)$ and both disappear. At the tip of the bi-stability region, where its upper and lower boundary meet, the system has a cusp point.

Different transects of the $p-g$ space keeping one of the parameters fixed show different $\rho_{\mbox{\tiny{C}}}^*$ vs $p$ (or $\rho_{\mbox{\tiny{C}}}^*$ vs $g$) bifurcation diagrams with important ecological implications for the population dynamics. A bifurcation diagram $\rho_{\mbox{\tiny{C}}}^*$ vs $g$ with $p=0.01$ shows that the establishment probability controls a classical hysteresis loop between two species-coexistence states, $\rho_{\mbox{\tiny{C}}}^3(p)$ and $\rho_{\mbox{\tiny{C}}}^1(p)$ [Fig.~\ref{phase-diag}(b)]. A bifurcation diagram $\rho_{\mbox{\tiny{C}}}^*$ vs $p$ with $g=0.75$ above the cusp point [Fig.~\ref{phase-diag}(c)] shows an imperfect pitchfork bifurcation.  Here, the lower piece consists entirely of stable fixed points $\rho_{\mbox{\tiny{C}}}^1(p)$, while the upper piece shows a saddle-node bifurcation in which a stable and an unstable fixed point, $\rho_{\mbox{\tiny{C}}}^3(p)$ and $\rho_{\mbox{\tiny{C}}}^2(p)$, respectively, approach to each other as $p$ increases from $0$, until $p = p_c$ where they meet and disappear. Conversely, if $g$ is kept constant below the cusp point, a bifurcation diagram $\rho_{\mbox{\tiny{C}}}^*$ vs $p$ shows  a reversed imperfect pitchfork, as compared to the above-cusp transect discussed before, where the upper branch of the bi-stable connects with the branch in the mono-stable region (not shown).  Finally, at the cusp point the system shows a perfect pitchfork bifurcation.

Both imperfect pitchfork bifurcations include a saddle-node bifurcation at $p_c$ that defines a discontinuous transition in the density of competitors and a {\it half} hysteresis loop with important ecological consequences (cusp catastrophe). If we set $p=0$ and $g=0.75$ [above the cusp point, Fig.~\ref{phase-diag}(c)] and start the system from the absorbing state $\rho_{\mbox{\tiny{C}}}^*=1$ corresponding to the dominance of species $C$, the system follows the upper branch $\rho_{\mbox{\tiny{C}}}^3(p)$ as $p$ increases, and undergoes a sharp transition at $p_c$ where the stationary density of species $C$ jumps from $\rho_{\mbox{\tiny{C}}}^3(p_c) \simeq 0.755$ to a lower value $\rho_{\mbox{\tiny{C}}}^1(p_c) \simeq 0.337$, and then increases until it reaches the value $\rho_{\mbox{\tiny{C}}}^1(1)=2/3$ at $p=1$.  However, the reverse path from $p=1$ to $p=0$ is always along the stable branch $\rho_{\mbox{\tiny{C}}}^1(p)$ until the point $\rho_{\mbox{\tiny{C}}}^1(0)=0$, corresponding to $D$--dominance.  Thus, once the system falls into the lower branch it can never reach a state with the dominance of $C$, and not even a density $\rho_{\mbox{\tiny{C}}}^*$ larger than $1/(1+g)$. From an ecological point of view, this half loop represents and even more dangerous transition for population persistence than the usual hysteresis loops like the one in Fig.~\ref{phase-diag}(b), since the initial $C$--dominance can never be recovered once the system overcomes the threshold $p_c$.

\end{enumerate}

In summary, when external noise is added to the system, two new phases of species coexistence appears, one with a unique stable coexistence and the other with two stable coexistences. The stationary densities in both phases vary with the germination and immigration probabilities $g$ and $p$.

\section{Monte Carlo results}
\label{Monte-Carlo}

\subsection{One-dimensional lattice}
\label{sub:1dMC}
The pair approximation developed in the previous sections provides a qualitative description of the model dynamics and is a good first approach to determine the dynamical regimes present in our model. However, the PA assumes that the system is infinitely large and takes into account only nearest-neighbor correlations. Hence, finite-size fluctuations and longer-distance correlations are neglected. To test the validity of the PA assumptions we present here results from extensive Monte Carlo (MC) simulations of the individual-level stochastic dynamics described in section \ref{model}. We started the simulations with a one-dimensional regular lattice in which each site is occupied with a competitor or a disperser with probabilities $\rho_{\mbox{\tiny{C}}}(0)$ and $1-\rho_{\mbox{\tiny{C}}}(0)$, respectively. Then, we ran the stochastic dynamics of Fig.~\ref{fig:model} until the system reached the stationary state.

\begin{figure}
  \includegraphics[width=0.495\textwidth]{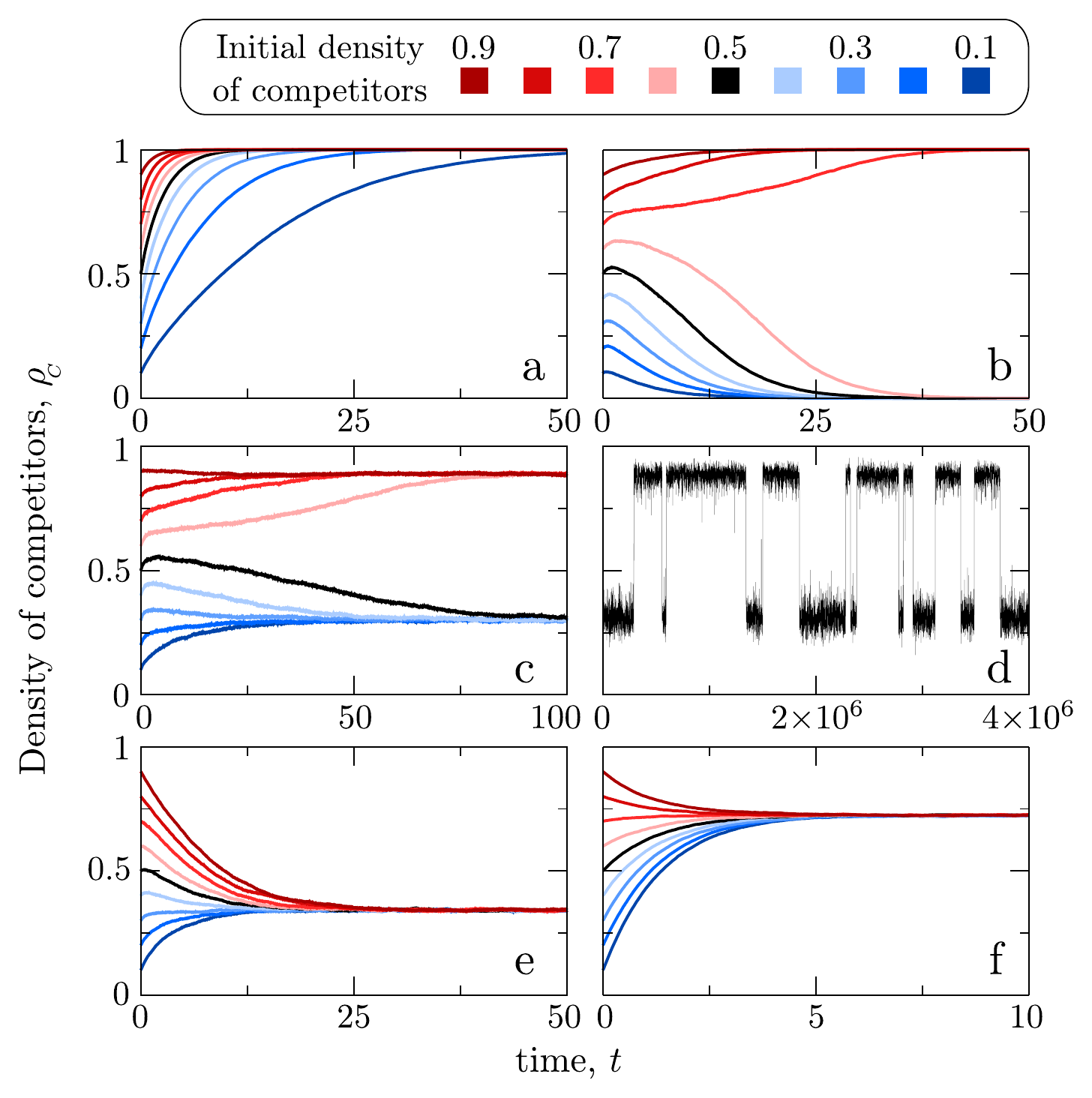}
  \caption{Single realizations of the stochastic particle dynamics starting at different initial densities of competitors (color code as indicated in the legend).  $N=10^5$ except in panel d). Note the different time scales in each panel. a) No immigration ($p=0$) and $g=0.15$ (competitor mono-stable dominance). b) No immigration and $g=0.75$ (bi-stable dominance).  c) $p=0.035$ and $g=0.7$ (bi-stable coexistence). d) same parameter values than e) but smaller system size, $N=10^3$, allow finite size fluctuations to induce transitions between the two coexistence points. e) $p=0.1$ and $g=0.9$ (mono-stable coexistence with more dispersers than competitors). f) $p=0.5$ and $g=0.5$ (mono-stable coexistence with more competitors than dispersers).}
  \label{fig:trajectories}
\end{figure}

In the no-immigration limit ($p=0$), all realizations eventually reach one of the two possible absorbing states [Fig.~\ref{fig:trajectories}(a) and (b)], confirming the species exclusion predicted by the PA calculations and numerical simulations in \citep{Durrett-1998}. Also in agreement with PA predictions, the system undergoes a transition from competitor dominance to bi-stable dominance as the establishment probability $g$ increases (note that in Fig. ~\ref{fig:phase} dispersers are always excluded for low $g$ but the dominant species depends on the initial condition for large $g$). However, MC simulations place the onset of bi-stability at $g(p=0) = 0.303$, whereas PA calculations give an estimated value $g = 1/2$.  To obtain the MC transition point to the bi-stable phase we ran spreading experiments on a lattice with $N=10^5$ and an initial condition consisting on all-dispersers with only a triplet of competitors located in the central sites of the lattice. Starting from this initial condition and using a resolution in $g$ of $10^{-3}$, we found that the system undergoes a transition from disperser dominance to competitor dominance at $g(p=0)\simeq 0.303$. 

A better analytical estimation of $g(p=0)$ can be obtained by noticing that, for a very small initial density of competitors $\rho_C(0) \ll 1$, the dynamics is akin to that of the contact process for disease spreading \citep{Durrett-1998} where competitors and dispersers are considered as infected and susceptible individuals, respectively.  In this case, the gain and loss transition probabilities Eqs. ~(\ref{W-}) and (\ref{W+}) are reduced to $\rho_C$ and $g \rho_C$,  respectively,  which corresponds to a contact process where infected individuals transmit the disease to a nearest-neighbor at rate $1/dt$ and they recover at rate $g/dt$.  Then, the disease-free state (dispersers dominance) looses stability when the ratio $1/g$ overcomes the transition value $3.2978$ for a one-dimensional system \citep{Marro1999},  i.e.,  at $g \simeq 0.3032$, which is in very good agreement with our MC simulations.

\begin{figure}
  \includegraphics[width=0.47\textwidth]{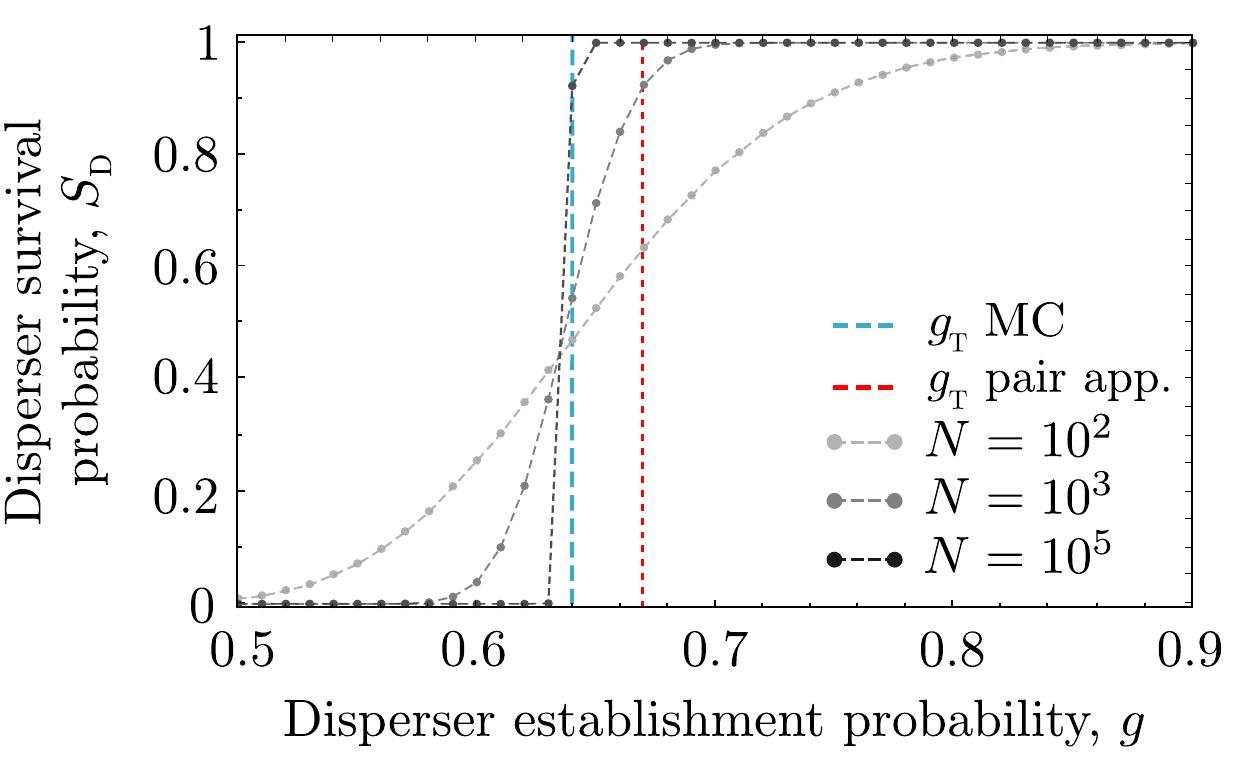}
  \caption{Disperser survival probability, $S_{\mbox{\tiny{D}}}$, as a function of disperser establishment probability calculated from Monte Carlo simulations in different system sizes (gray symbols and dashed lines). For each system size and establishment probability, we ran $1000$ independent model realizations and calculated disperser survival probability as the fraction of realizations in which dispersers excluded competitors (i.e., the system reached the absorbing state $\rho_{\mbox{\tiny{C}}}=0$). The vertical red dashed line shows the PA estimated value of $g_{\mbox{\tiny{T}}}$ and the vertical blue dashed line the MC estimated value of $g_{\mbox{\tiny{T}}}$ for $N=10^5$.}
  \label{fig:systemsize}
\end{figure}

On the other hand, for a fixed initial density of competitors, the establishment probability $g$ controls a transition from competitor dominance at low $g$ to disperser dominance at high $g$. This transition is continuous for small systems, indicating that finite size fluctuations can cause the extinction of the species that is favored by a specific value of $g$. As system size increases, however, the transition gets sharper, probably becoming discontinuous for infinite systems (Fig.~\ref{fig:systemsize}). However, the transition point from competitor to disperser dominance $g_{\mbox{\tiny{T}}}$, which we defined as the lowest value of the establishment probability for which dispersers are more likely to survive than competitors, $S_{\mbox{\tiny{D}}}>1/2$, is almost independent of system size. The pair approximation analysis also predicts this transition but it overestimates the value of $g_{\mbox{\tiny{T}}}$ except for very high initial densities of competitors (note the difference between the white dashed line and the red-blue frontier in Fig.~\ref{fig:phase}). For the specific case of $\rho(0)_{\mbox{\tiny{C}}}=0.5$ shown in Fig.~\ref{fig:systemsize}, MC simulations give an estimated $g_{\mbox{\tiny{T}}}\approx 0.64$ (for $N=10^5$) and the PA predicts  $g_{\mbox{\tiny{T}}}\approx 0.68$.

\begin{figure}
  \includegraphics[width=0.46\textwidth]{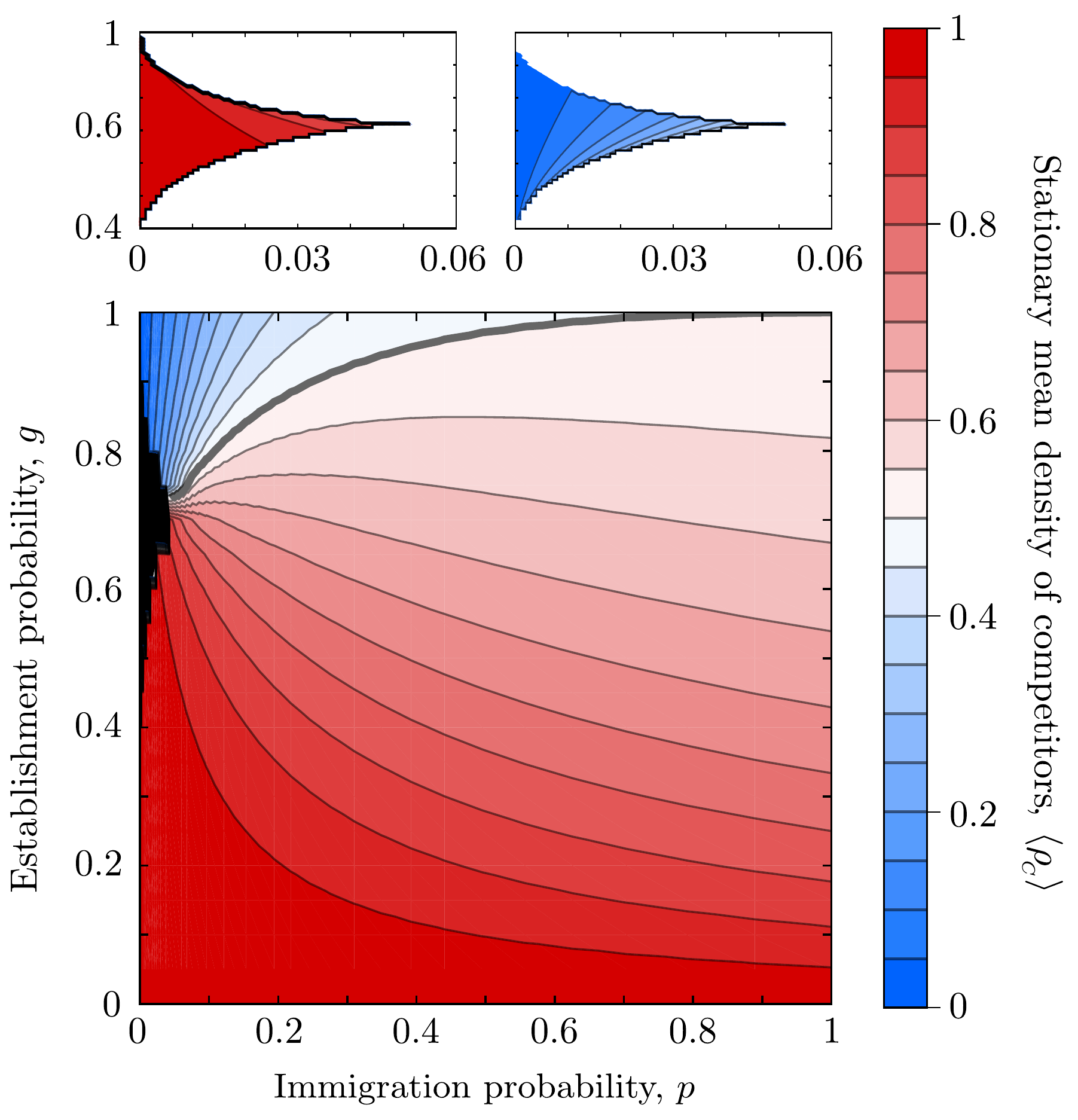}
  \caption{Mean density of competitors in the stationary state, $\langle \rho_{\mbox{\tiny{C}}}\rangle$, obtained from Monte Carlo simulations with a system size $N=10^5$. Averages are taken both in time (once the system is in the stationary state $10^5>t>5\times 10^5)$ and over $50$ independent realizations. The thicker contour indicates $\langle \rho_{\mbox{\tiny{C}}}\rangle=0.5$. The top panels show $\langle \rho_{\mbox{\tiny{C}}}\rangle$ obtained in the bi-stable phase (black region of the main panel) for two different initial conditions [$\rho_{\mbox{\tiny{C}}}(0)=0.99$ (left) and $\rho_{\mbox{\tiny{C}}}(0)=0.01$ (right)].}
  \label{fig:MC-full}
\end{figure}

MC simulations qualitatively corroborate PA predictions when immigration is allowed ($p\neq 0$) too. First, we tested the existence of bi-stable coexistence in MC simulations. For various values of $p$ and $g$, we ran simulations with two different initial densities of competitors, $\rho_{\mbox{\tiny{C}}}(0)=0.99$ and $\rho_{\mbox{\tiny{C}}}(0)=0.01$, and calculated the mean density of competitors in the steady state for each of them. Because we used large systems, we do not observe noise-induced transitions between steady states within each realization, which allowed us to use the difference between the mean density of competitors for each initial condition as a test for bi-stability (see Fig.~\ref{fig:trajectories}(c) and \ref{fig:trajectories}(d) for a comparison of model realizations in the bi-stable phase using different system sizes).  Finally, because the system loses bi-stability through a catastrophic transition, this difference jumps abruptly from a non-zero to a zero value [see Fig.~\ref{phase-diag}(c)]. We placed the frontier of the bi-stable phase at the values of  $p$ and $g$ for which the difference in the stationary mean density of competitors reached from $\rho_{\mbox{\tiny{C}}}(0)=0.99$ and $\rho_{\mbox{\tiny{C}}}(0)=0.01$ is smaller than $10^{-5}$. Using this definition,  MC simulations confirm that the model may show bi-stable coexistence if immigration is weak ($0<p\lesssim 0.06$). That is, species reach a steady state of coexistence in which the density of each species depends on the initial conditions [light gray region in Fig.~\ref{phase-diag}(a)]. When immigration becomes more frequent ($p$ increases), bi-stability is lost but species still coexist at varying population proportions. For large establishment probability and weak immigration probability, dispersers dominate the population [blue region on the top-left corner of Fig.~\ref{fig:MC-full}; Fig.~\ref{fig:trajectories}(e)] However, when immigration probability increases and establishment probability decreases, competitors take over [Fig.~\ref{fig:trajectories}(f); red region in Fig.~\ref{fig:MC-full}]. In the only-immigration limit ($p=1$) the dispersal component of the competition-dispersal tradeoff is immaterial and the frequency of each species in the mix is determined by its probability to establish following an immigration event. Species therefore coexist at a mean density of competitors $\langle\rho_{\mbox{\tiny{C}}}\rangle = 1/(1+g)$,  as predicted by the PA approximation.

\subsection{Two-dimensional lattice}\label{sec:MC2D}
To test the generality of our one-dimensional results, we conducted numerical simulations in a two-dimensional regular lattice of lateral length $\ell$ and using periodic boundary conditions. In this case, competitors can invade one of their four nearest neighbors upon recruitment, and dispersers any other lattice site. All the other model components follow the one-dimensional dynamics introduced in Section \ref{model}. 

\begin{figure}
  \includegraphics[width=0.46\textwidth]{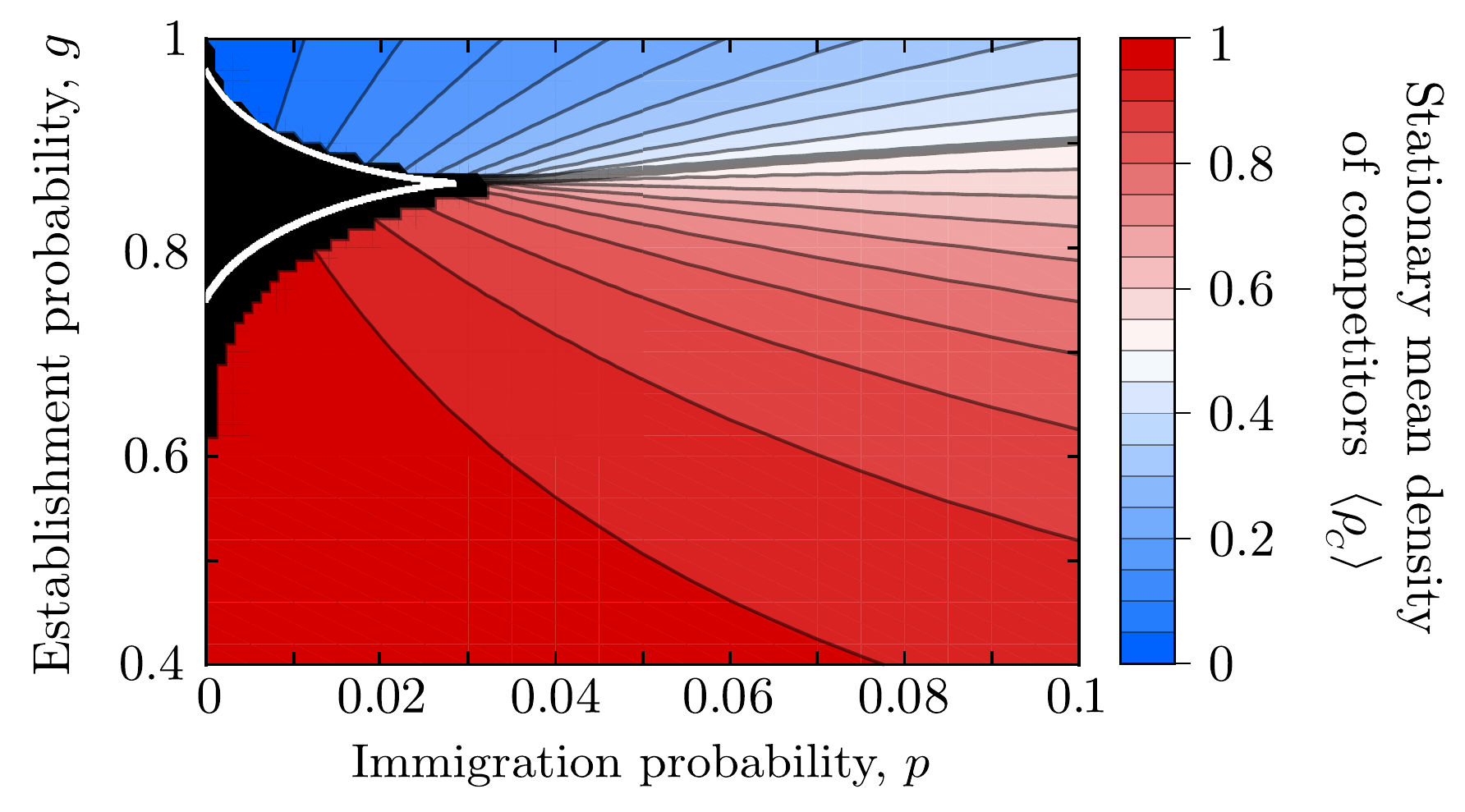}
  \caption{Mean density of competitors in the stationary state, $\langle \rho_{\mbox{\tiny{C}}}\rangle$, obtained from Monte Carlo simulations in a two-dimensional lattice with lateral length $\ell = 200$. Note the different scale in the axes compared to Fig.\,\ref{fig:MC-full}. Averages are taken both in time (once the system is in the stationary state $(10^5>t>5\times 10^5)$ and over $50$ independent realizations. The thicker contour indicates $\langle \rho_{\mbox{\tiny{C}}}\rangle=0.5$. The black region indicates the bi-stable phase, which was determined by identifying the combination of $p$, $g$ values that lead to different stationary states when starting simulations with different initial conditions [$\rho_{\mbox{\tiny{C}}}(0)=0.99$ (left) and $\rho_{\mbox{\tiny{C}}}(0)=0.01$ (right)]. The white line limits the region of bi-stability predicted by the pair approximation (see Appendix \ref{PA-d} for full derivation).}
  \label{fig:2DMCl}
\end{figure}

These numerical simulations confirm that the behavior of the model in one-dimensional systems can be qualitatively extended to two dimensions. In the no-immigration limit ($p=0$) we find that competitors dominate at low disperser establishment probability $g$, while for high values of $g$ a bi-stable dominance is observed. Starting with very low densities of competitors, our model without immigration can be mapped to a contact process following the same rationale discussed in section \ref{sub:1dMC} for the one-dimensional case. We estimate a transition from competitor dominance to bi-stable dominance at $g\approx 0.606$, in good agreement with results for the contact process in $2D$ regular lattices \cite{Marro1999}. When immigration is allowed, the system reaches a stationary state of mono-stable species coexistence in most of the $(p,g)$ parameter space (Fig.\,\ref{fig:2DMCl}), but bi-stable coexistence is also possible for low $p$ and large $g$ (black region in Fig.\,\ref{fig:2DMCl}).

Implementing a PA approach akin to that developed in section~\ref{mean-field} for the model in a one-dimensional lattice, we obtained the following approximate equations for the evolution of the system in two-dimensional regular lattices (see Appendix \ref{PA-d} for the general derivation of the PA equations in $d$-dimensional lattices):
\begin{subequations}
  \begin{alignat}{2}
    \label{drhoSdt-1-2D}
    \frac{d \rho_{\mbox{\tiny{C}}}}{dt} &= (1-p) \left[\rho_{\mbox{\tiny{CD}}} - g \,
      \rho_{\mbox{\tiny{C}}}(1-\rho_{\mbox{\tiny{C}}})\right] + p \left[ 1-(1+g)\rho_{\mbox{\tiny{C}}} \right], \\
    \label{drhoSLdt-1-2D}
    \frac{d \rho_{\mbox{\tiny{CD}}}}{dt} &= (1-p) \left[2 g(1-\rho_{\mbox{\tiny{C}}})(\rho_{\mbox{\tiny{C}}}-2\rho_{\mbox{\tiny{CD}}}) +  \rho_{\mbox{\tiny{CD}}}- \frac{3 \rho_{\mbox{\tiny{CD}}}^2}{1-\rho_{\mbox{\tiny{C}}}} \right] \nonumber \\
    & + 2p \left[1-(1-g) \rho_{\mbox{\tiny{C}}}-2(1+g) \rho_{\mbox{\tiny{CD}}} \right],
  \end{alignat}
  \label{rhoS-rhoSL-2D}
\end{subequations}   
which further confirms the existence of the cusp catastrophe found in MC simulations in two-dimensional lattices (white line in Fig. \ref{fig:2DMCl}).

\section{Summary and conclusions}
\label{discussion}
In this article we introduced a modified version of the noisy voter model to investigate the competition between two species to colonize a territory. The two species differ in their dispersion range and their ability to invade sites. Specifically, one of the species is able to disperse offspring to any lattice site but they have a reduced probability of establishing upon dispersal $g\leq1$,  whereas the other species' offspring can only reach nearest-neighbor sites but replace non-specific residents with probability $1$. The model also accounts for immigration, which causes the spontaneous replacement of individuals belonging to one species by individuals of the other species. The time scale of the immigration process compared to that of recruitment is controlled by a parameter $p$ that measures how often immigration events take place compared to reproduction. We conducted an intensive numerical and analytical study of the model in one and two-dimensional lattices.

 The different ways in which each of the two species balances a competition-dispersal tradeoff, together with the effect of immigration, lead to a rich variety of possible competition outcomes that depend on the relation between the germination probability $g$ and the frequency of immigration events $p$. For the one-dimensional system, in the absence of immigration ($p=0$) the system eventually reaches an absorbing state in which one of the species is excluded and the other occupies the entire territory. For $g \lesssim 0.303$, the ability of dispersers to invade distant sites is not enough to compensate its low chances of establishing upon dispersal, resulting in a dominance of competitors regardless of the initial composition of the population. At $g \approx 0.303$, however, the system undergoes a transition to a bi-stable phase in which either of the species can outcompete the other depending on the initial composition of the population. PA calculations qualitatively recapitulate these results, but overestimate the onset of bi-stability ($g_T=1/2$). In two-dimensional systems, the transition from mono-stable to bi-stable dominance occurs for larger values of $g$ than in one-dimensional $g \approx 0.606$ lattices, and the PA approximation also overestimates the onset of bi-stability ($g_T=3/4$). 
 
 This bi-stable dominance observed for large values of $g$ when $p=0$ is extended to a situation of bi-stable coexistence of both species when immigration is allowed but weak ($0<p<p_c(g)$). In this bi-stable phase, there are two possible stable steady states that correspond to two different case scenarios of asymmetric coexistence in which one species has a larger stationary density than the other. For values of $p$ larger than a threshold $p_c(g)$ the bi-stability is lost and replaced by a mono-stable coexistence, which is also observed for $g \lesssim 0.303$(1D) or $g \lesssim 0.606$(2D) if $p\neq 0$. These results are also qualitatively supported by a pair approximation both in one and two-dimensional lattices. In this region of mono-stable coexistence, competitors dominate for most $(p, g)$ parameter combinations because immigration is species independent and competitors have a larger establishment probability. In the limit $p=1$ the dynamics reduces to a sequence of immigration events and both species coexist at a frequency that is determined by dispersers establishment probability $g$, $\langle\rho_{\mbox{\tiny{C}}}\rangle = 1/(1+g)$.
 
At $p_c(g)$, the transition between the bi-stable and the mono-stable phase is abrupt, and gives rise to an irreversible hysteresis loop in the stationary density of competitors. In this dynamical regime and starting from a stationary population with high abundance of competitors, the density of competitors decreases as $p$ increases.  When $p$ overcomes the threshold $p_c(g)$, the system becomes mono-stable and the population jumps abruptly to a much lower density of competitors, which are now less abundant than dispersers. Mathematically, this means that the system jumps from the upper to the lower stable branch of the stable equilibria. Following this abrupt transition, the system always moves along the lower branch as $p$ is varied, and it can never jump back to the upper branch. As a consequence, if $p$ goes back to very low values, the population moves towards a high dominance of dispersers and it is not possible for competitors to take over and become more abundant than dispersers any more. In a real ecological system, this irreversible path might bring important consequences for the persistence of species with short colonization ranges, especially in time-varying environments with seasonal or fluctuating dispersal rates \citep{King1983,Mangan2002,Nathan2005,Williams2013}. Future work should focus on better understanding this dynamics and its potential consequences for biodiversity maintenance.

Our results also suggest other possible directions for future work. We have shown that in voter-like models (also termed replacement models in the ecological literature \citep{Yu2001,Calcagno2006}) a competition-dispersal tradeoff only allows long-term species coexistence when coupled to immigration. Our model, however, provides an analytically tractable framework to investigate whether spatial heterogeneity, either in the establishment probability or the dispersal range, could promote species coexistence even without immigration \citep{Mordecai2016}.  Moreover, we have considered that dispersers have an infinity dispersal range, which facilitated the analytical treatment of the model. Relaxing this assumption to allow large but finite dispersal ranges \cite{Durrett-1998} could provide deeper insights on the role that competition-dispersal tradeoffs play in determining the conditions for species coexistence in more realistic communities \citep{Higgins2002}. Finally, it would be interesting to extend our model to consider species-specific immigration probabilities. Our results show a large region of the parameter space in which competitors dominate over dispersers, which is likely due to the fact that in the high immigration limit individuals from each species arrive at the same rate but have different establishment probabilities. Favoring the arrival of dispersers over competitors, assuming that both intra-patch and inter-patch dispersal are mediated by the same mechanism, would probably balance this difference. From a theoretical point of view, it would be worthwhile to explore other possible dynamics that could lead to coexistence in voter-like models, such as those implemented in the non-linear voter models for social dynamics \cite{Vazquez-2008-b,Castellano-2009, Martinez-Garcia2012}. 

\begin{acknowledgments}
We thank two anonymous reviewers for their constructive comments and suggesting the pair-approximation analysis of the two-dimensional model. RMG acknowledges financial support from the Simons Foundation, Instituto Serrapilheira through grant Serra-1911-31200, and FAPESP through the ICTP-SAIFR grant 2016/01343-7 and Programa Jovens Pesquisadores em Centros Emergentes 2019/24433-0, 2019/05523-8. CL acknowledges financial support from the Spanish State Research Agency through the María de Maeztu Program for Units of Excellence in R\&D (MDM-2017-0711). FV acknowledges the University of the Balearic Islands for a grant  under program {\it Visiting of lecturers} and financial support from CONICET (Grant No. PIP 0443/2014) and from Agencia Nacional de Promoci\'on Cient\'ifica y Tecnol\'ogica (Grant No. PICT 2016 Nro 201 0215).
\end{acknowledgments}

%

\vspace{1.5cm}
\onecolumngrid
\begin{center}
\rule {17cm}{0.04cm}
\end{center}
\appendix
\section{Stability analysis for the $p=0$ case} 

\label{stability}

We start by analyzing the stability of the trivial fixed point $(0,0)$.  For that, we linearize Eqs.~(\ref{drhoSdt}) and (\ref{drhoSLdt}) around $(0,0)$ and obtain the following system written in matrix representation:
\begin{eqnarray*}
 \frac{d \vec{\epsilon}}{dt} = {\bf A} \, \vec{\epsilon},
\end{eqnarray*}
with  
\begin{eqnarray*}
 {\bf A} \equiv \left( {\begin{array}{cc}
 -g & 1 \\
 2g & -4 g \\
\end{array} } \right)
\end{eqnarray*}
and $\vec{\epsilon} \equiv (\epsilon_1,\epsilon_2)$, where the components of $\vec{\epsilon}$ are small independent perturbations of the fixed $(0,0)$, i.e., $\rho_{\mbox{\tiny C}}=\epsilon_1$ and $\rho_{\mbox{\tiny{CD}}}=\epsilon_2$.  The eigenvalues of ${\bf A}$ are 
\begin{eqnarray*}
  \lambda_{\pm} = \frac{1}{2} \left[ -5g \pm \sqrt{g(9g+8)} \right].
\end{eqnarray*}
Then, $(0,0)$ is a stable fixed point along the direction associated to $\lambda_-$ for all values of $g$ in $[0,1]$, while it is stable for $g>1/2$ and unstable for $g<1/2$ along the direction associated to $\lambda_+$.  Therefore, $(0,0)$ is stable for $g>1/2$ and a saddle point for $g<1/2$.

A similar stability analysis around the coexistence fixed point $\vec{\rho_{co}}^* = \left(\frac{2g-1}{g}, \frac{(1-g)(2g-1)}{g}\right)$ leads to the eigenvalues  
\begin{eqnarray*}
  \lambda_{\pm} = \frac{1}{2} \left[ -(2+g) \pm \sqrt{-4+28g-15g^2} \right].
\end{eqnarray*}
Then, for $0.155 < g < 1.71$ the eigenvalues $\lambda_{\pm}$ are real,while outside this interval are complex with a negative real part. Given that the physically possible values of $g$ are in the interval $[0,1]$, we find that $\vec{\rho_{co}}$ is a stable spiral fixed point for $0 \le g < 0.155$.  Besides, for $0.155 < g < 1/2$ both $\lambda_{\pm}$ are real and negative, and so $\vec{\rho_{co}}$ is stable, while for $1/2 < g \le 1$ is $\lambda_-<0$ and $\lambda_+>0$, and so $\vec{\rho_{co}}$ is a saddle point.  Finally, the fixed point $(1,0)$ is stable for all $g \in [0,1]$.  

In summary, $\vec{\rho_{co}}^*$ and $(1,0)$ are stable for $0 \le g <1/2$, while $(0,0)$ and $(1,0)$ are stable for $1/2 < g \le 1$. We have checked numerically that for $0 \le g < 1/2$, starting from a ``non-physical'' initial condition with $\rho_{\mbox{\tiny{CD}}}(0)<0$, the evolution of $\rho_{\mbox{\tiny{C}}}(t)$ and $\rho_{\mbox{\tiny{CD}}}(t)$ exhibit dumped oscillations in its approach to the fixed point $\vec{\rho_{co}}^*$, while for any real physical initial condition $\rho_{\mbox{\tiny{CD}}}(0)>0$ and $\rho_{\mbox{\tiny{C}}}(0)>0$ the evolution is towards the stable fixed point $(1,0)$.  Also, for $1/2 < g \le 1$ the evolution is towards $(0,0)$ or $(1,0)$ depending on the initial condition, as we explain in section~\ref{recruitment}. \\

\section{Fixed points for the $p>0$ case}
\label{fixed-points}

In this section we show how to obtain numerical estimates of the fixed points of Eqs.~(\ref{rhoS-rhoSL}) for $p>0$.  We set to zero the left-hand side of Eqs.~(\ref{rhoS-rhoSL}) and solve for $\rho_{\mbox{\tiny{C}}}$ from Eq.~(\ref{drhoSdt-1}), which leads to the following relation between the stationary values of $\rho_{\mbox{\tiny{CD}}}$ and $\rho_{\mbox{\tiny{C}}}$: 
\begin{eqnarray*}
  \rho_{\mbox{\tiny{CD}}}^* = g\, \rho_{\mbox{\tiny{C}}}^* (1 -\rho_{\mbox{\tiny{C}}}^*) - \frac{p \left[ 1 - (1 + g) \rho_{\mbox{\tiny{C}}}^* \right]}{1 - p}.
\end{eqnarray*}
We now plug this expression for $\rho_{\mbox{\tiny{CD}}}^*$ into Eq.~(\ref{drhoSLdt-1}) at the stationary state and obtain the condition
\begin{eqnarray}\label{f}
  f(\rho_{\mbox{\tiny{C}}}^*)= A + B \rho_{\mbox{\tiny{C}}}^*+ C (\rho_{\mbox{\tiny{C}}}^*)^2+ D(\rho_{\mbox{\tiny{C}}}^*)^3+ E (\rho_{\mbox{\tiny{C}}}^*)^4 =0,
\end{eqnarray}
assuming that $\rho_{\mbox{\tiny{C}}}^* \neq 1$, where the coefficients $A,B,C,D$ and $E$ are given by
\begin{eqnarray*}
 A &=& \frac{2 (1 + 2g) p}{1 - p}, \\
 B &=& \frac{2 \left[-2 g^2 - 2 p +  g (1 - 7 p + 2 p^2)\right]}{1 - p}, \\
 C &= & \frac{2 \left[ p - g^2 (-5 + 4 p) - g (2 - 9 p + 5 p^2) \right]}{1 - p}, \\
 D &=& 2 g \left[ 1 + 2 g (p-2) - 3 p \right], \\
 E &=& 2 g^2 (1 - p). 
\end{eqnarray*}
The four roots of the polynomial $f(\rho_{\mbox{\tiny{C}}}^*)$ from Eq.~(\ref{f}) correspond to the fixed points of Eqs.~(\ref{rhoS-rhoSL}).  We numerically found the roots of $f(\rho_{\mbox{\tiny{C}}}^*)$ with an approximate error of $10^{-7}$. Depending on the values of $p$ and $g$, only one root lays in the physical interval $\rho_{\mbox{\tiny{C}}}^* \in [0,1]$ for $0 \le g < 1/2$, while one or three roots are in $[0,1]$ for $1/2 < g \le 1$, as we describe in section~\ref{immigration}.

\section{Pair approximation in $d$--dimensional lattices}
\label{PA-d}

In this section we derive rate equations for the evolution of $\rho_{\mbox{\tiny{C}}}$ and $\rho_{\mbox{\tiny{CD}}}$ in lattices of dimension $d$.  We follow a PA approach that is akin to that developed in section~\ref{mean-field} for $1D$ lattices, and that we now extend to a generic dimension $d \ge 1$, where each lattice site has $z \equiv 2^d$ nearest neighbors (NNs).  In a single time step $dt=1/N$, a site $i$ with state $\sigma_i = \{C,D\}$ is chosen at random.  Then, either an immigration event takes place with probability $p$ or a recruitment event happens with the complementary probability $1-p$.

\bigskip
 
{\bf Immigration:}
\begin{enumerate}
 \item With probability $\rho_{\mbox{\tiny{D}}}=1-\rho_{\mbox{\tiny{C}}}$, site $i$ is in state $\sigma_i=D$, and then its state is switched with probability $1$ ($\sigma_i=D \to \sigma_i=C$).  This leads to a change $\Delta \rho_{\mbox{\tiny{C}}}=1/N$ 
in the density of $C$--sites and to a net change 
\begin{equation}
	\Delta \rho_{\mbox{\tiny{CD}}}=\frac{2(z-2 \, n_{\mbox{\tiny{DC}}})}{zN}
	\label{Delta-CD}
\end{equation}
 in the density of $CD$--pairs, where $n_{\mbox{\tiny{DC}}}$ is the number of NNs of site $i$ that are in state $C$ ($0\le n_{\mbox{\tiny{DC}}} \le z$), i.e., $CD$--pairs centered at $\sigma_i=D$.  That is, if initially there are $n_{\mbox{\tiny{DC}}}$ $CD$--pairs around $\sigma_i=D$, the number of $CD$--pairs after $i$ switches state is $z-n_{\mbox{\tiny{DC}}}$  (the initial $DD$--pairs become $CD$--pairs).  This gives a net change $z-2 \, n_{\mbox{\tiny{DC}}}$ in the total number of $CD$--pairs in the system, which becomes expression Eq.~(\ref{Delta-CD}) for $\Delta \rho_{\mbox{\tiny{CD}}}$ when we normalize by the total number of NNs pairs $zN/2$.  

\item With probability $\rho_{\mbox{\tiny{C}}}$, site $i$ is in state $\sigma_i = C$, and thus it switches state with probability $g$ ($\sigma_i=C \to \sigma_i=D$), leading to the changes $\Delta \rho_{\mbox{\tiny{C}}}=-1/N$ and 
\begin{equation}
	\Delta \rho_{\mbox{\tiny{CD}}}=\frac{2(z-2\,n_{\mbox{\tiny{CD}}})}{zN},
	\label{Delta-CD-1}
\end{equation}
where $n_{\mbox{\tiny{CD}}}$ is the initial number of $CD$--pairs centered at $\sigma_i=C$.

\end{enumerate} 

Assembling these factors, the average change of $\rho_{\mbox{\tiny{C}}}$ in a time step can be calculated as  
\begin{eqnarray}
	\frac{d \rho_{\mbox{\tiny{C}}}}{dt} = \frac{1}{1/N} \left( \rho_{\mbox{\tiny{D}}} \frac{1}{N} - g \, \rho_{\mbox{\tiny{C}}} \frac{1}{N} \right) = 1- (1+g) \rho_{\mbox{\tiny{C}}},
	\label{drCdt-imm}
\end{eqnarray}
while the average change of $\rho_{\mbox{\tiny{CD}}}$ is given by
\begin{eqnarray}
	\frac{d \rho_{\mbox{\tiny{CD}}}}{dt} = \frac{\rho_{\mbox{\tiny{D}}}}{1/N} \sum_{n_{\mbox{\tiny{DC}}}=0}^z B\left( n_{\mbox{\tiny{DC}}},z;P_{\mbox{\tiny{C$|$D}}} \right) \frac{2(z-2 \, n_{\mbox{\tiny{DC}}})}{zN} + 
	\frac{g \, \rho_{\mbox{\tiny{C}}}}{1/N} \sum_{n_{\mbox{\tiny{CD}}}=0}^z B\left( n_{\mbox{\tiny{CD}}},z;P_{\mbox{\tiny{D$|$C}}} \right) \frac{2(z-2 \, n_{\mbox{\tiny{CD}}})}{zN}, 
	\label{drhoCDdt-z}
\end{eqnarray}
where $B\left( n_{\mbox{\tiny{DC}}},z;P_{\mbox{\tiny{C$|$D}}} \right)$ is the probability that there are $n_{\mbox{\tiny{DC}}}$ $CD$--pairs around a $D$--site that has $z$ NNs, and $P_{\mbox{\tiny{C$|$D}}}$ is the conditional probability that a NN of a $D$--site is a $C$--site, and similarly for $B\left( n_{\mbox{\tiny{CD}}},z;P_{\mbox{\tiny{D$|$C}}} \right)$.  If we assume that the states of second nearest neighbors are uncorrelated (pair approximation), $B$ becomes the Binomial distribution, with first moments
\begin{eqnarray}
	\langle n_{\mbox{\tiny{DC}}} \rangle = z \, P_{\mbox{\tiny{C$|$D}}} = \frac{z \, \rho_{\mbox{\tiny{CD}}}}{\rho_{\mbox{\tiny{D}}}} ~~~ \mbox{and} ~~~
	\langle n_{\mbox{\tiny{CD}}} \rangle = z \, P_{\mbox{\tiny{D$|$C}}} = \frac{z \, \rho_{\mbox{\tiny{CD}}}}{\rho_{\mbox{\tiny{C}}}},
	\label{moments}
\end{eqnarray}
where we have used expressions Eqs.~(\ref{conditional}) for the conditional probabilities $P_{\mbox{\tiny{C$|$D}}}$ and $P_{\mbox{\tiny{D$|$C}}}$.  Expanding Eq.~(\ref{drhoCDdt-z}) we obtain 
\begin{eqnarray}
	\frac{d \rho_{\mbox{\tiny{CD}}}}{dt} = \frac{2 \, \rho_{\mbox{\tiny{D}}}}{z} \left(  z - 2 \langle n_{\mbox{\tiny{DC}}} \rangle \right) + \frac{2 \, g \, \rho_{\mbox{\tiny{C}}}}{z} \left(  z - 2 \langle n_{\mbox{\tiny{CD}}} \rangle \right),
\end{eqnarray}
and replacing the expressions for the moments from Eq.~(\ref{moments}) we finally arrive at
\begin{eqnarray}
	\frac{d \rho_{\mbox{\tiny{CD}}}}{dt} = 2 \left[ 1-(1-g) \rho_{\mbox{\tiny{C}}} - 2(1+g) \rho_{\mbox{\tiny{CD}}} \right].
	\label{drCDdt-imm}
\end{eqnarray}

\bigskip 

{\bf Recruitment:} 

\begin{enumerate} 

\item Site $i$ is in state $\sigma_i=D$ with probability $\rho_{\mbox{\tiny{D}}}$, which then choses and invades a random site $j \ne i$ in state $\sigma_j=C$ with probability $g \, \rho_{\mbox{\tiny{C}}}$ ($D ... C \to D ... D$).  This leads to changes $\Delta \rho_{\mbox{\tiny{C}}}=-1/N$ and $\Delta \rho_{\mbox{\tiny{CD}}}$ as given by Eq.~(\ref{Delta-CD-1}).

\item Site $i$ is in state $\sigma_i=C$ with probability $\rho_{\mbox{\tiny{C}}}$, which then chooses and invades a random NN $D$--site $j$ with probability $P_{\mbox{\tiny{D$|$C}}}=\rho_{\mbox{\tiny{CD}}}/\rho_{\mbox{\tiny{C}}}$ ($CD \to CC$).  Then, the associated changes are $\Delta \rho_{\mbox{\tiny{C}}}=1/N$ and $\Delta \rho_{\mbox{\tiny{CD}}}=2[z-2(1+n'_{\mbox{\tiny{DC}}})]/(zN)$.  Here, the total number of $CD$--pairs centered at site $\sigma_j=D$ is $n_{\mbox{\tiny{DC}}}=1+n'_{\mbox{\tiny{DC}}}$, which is composed by the chosen $CD$--pair and $n'_{\mbox{\tiny{DC}}}$ $CD$--pairs over the other $z-1$ NNs of $j$.

\end{enumerate}

Combining these factors we obtain
\begin{eqnarray}
	\frac{d \rho_{\mbox{\tiny{C}}}}{dt} = \rho_{\mbox{\tiny{CD}}} - g \, \rho_{\mbox{\tiny{C}}} (1-\rho_{\mbox{\tiny{C}}}), ~~~ \mbox{and}
	\label{drCdt-rec}
\end{eqnarray}

\begin{eqnarray}
	\frac{d \rho_{\mbox{\tiny{CD}}}}{dt} &=& \frac{g \, \rho_{\mbox{\tiny{D}}} \, \rho_{\mbox{\tiny{C}}}}{1/N} \sum_{n_{\mbox{\tiny{CD}}}=0}^z B\left( n_{\mbox{\tiny{CD}}},z;P_{\mbox{\tiny{D$|$C}}} \right) \frac{2(z-2 \, n_{\mbox{\tiny{DC}}})}{zN} + 
	\frac{\rho_{\mbox{\tiny{CD}}}}{1/N} \sum_{n'_{\mbox{\tiny{DC}}}=0}^{z-1} B\left( n'_{\mbox{\tiny{DC}}},z-1;P_{\mbox{\tiny{C$|$D}}} \right) \frac{2[z-2 (1+\, n'_{\mbox{\tiny{DC}}})]}{zN}, \\
	&=& \frac{2g \, \rho_{\mbox{\tiny{D}}} \, \rho_{\mbox{\tiny{C}}}}{z} \left( z-2 \langle n_{\mbox{\tiny{CD}}} \rangle \right) + \frac{2 \rho_{\mbox{\tiny{CD}}}}{z} \left( z-2-2 \langle n'_{\mbox{\tiny{DC}}} \rangle \right),
\end{eqnarray}
which after replacing $\langle n_{\mbox{\tiny{CD}}} \rangle$ by Eq.~(\ref{moments}) and $\langle n'_{\mbox{\tiny{DC}}} \rangle$ by $(z-1)\rho_{\mbox{\tiny{CD}}}/\rho_{\mbox{\tiny{D}}}$ becomes
\begin{eqnarray}
	\frac{d \rho_{\mbox{\tiny{CD}}}}{dt} = 2 g (1-\rho_{\mbox{\tiny{C}}}) (\rho_{\mbox{\tiny{C}}}-2\rho_{\mbox{\tiny{CD}}}) + \frac{2 \rho_{\mbox{\tiny{CD}}}}{z} \left[z-2-\frac{2(z-1) \rho_{\mbox{\tiny{CD}}}}{1-\rho_{\mbox{\tiny{C}}}} \right]. 
	\label{drCDdt-rec}
\end{eqnarray}

Finally, combining Eqs.~(\ref{drCdt-imm}) and (\ref{drCDdt-imm}) for immigration with Eqs.~(\ref{drCdt-rec}) and (\ref{drCDdt-rec}) for recruitment, we obtain the following set of approximate rate equations for the evolution of the densities of $C$--sites and $CD$--pairs in lattices of dimension $d$:
\begin{subequations}
  \begin{alignat}{2}
	\frac{d \rho_{\mbox{\tiny{C}}}}{dt} &= (1-p) \left[ \rho_{\mbox{\tiny{CD}}} - g \, \rho_{\mbox{\tiny{C}}} (1-\rho_{\mbox{\tiny{C}}}) \right]   + p \left[ 1- (1+g) \rho_{\mbox{\tiny{C}}} \right], \\
	\frac{d \rho_{\mbox{\tiny{CD}}}}{dt} &= 2 (1-p)  \Bigg\{ g (1-\rho_{\mbox{\tiny{C}}}) (\rho_{\mbox{\tiny{C}}}-2\rho_{\mbox{\tiny{CD}}}) + \frac{ \rho_{\mbox{\tiny{CD}}}}{z} \left[z-2-\frac{2(z-1) \rho_{\mbox{\tiny{CD}}}}{1-\rho_{\mbox{\tiny{C}}}} \right] \Bigg\} + 2p \left[ 1-(1-g) \rho_{\mbox{\tiny{C}}} - 2(1+g) \rho_{\mbox{\tiny{CD}}} \right]. 
	  \end{alignat}
	\label{drC-CD-dt}
\end{subequations}   
We can check that Eqs.~(\ref{drC-CD-dt}) become Eqs.~(\ref{rhoS-rhoSL}) for $z=2$ ($d=1$) and Eqs.~(\ref{rhoS-rhoSL-2D}) for $z=4$ ($d=2$).

\end{document}